\documentclass[aoas,MSNbibl,nameyear,dvips]{arximspdf}
\usepackage{graphicx}

\doi{10.1214/11-AOAS484}
\volume{5}
\issue{3}
\pubyear{2011}
\firstpage{2231}
\lastpage{2263}

\makeatletter

\def\bsuffix #1{#1}

\newcommand{\eqref}[1]{(\ref{#1})}

\newcommand{\gs}{\gtrsim}
\newcommand{\ls}{\lesssim}

\newcommand{\thetab}{\bolds\theta}
\newcommand{\bell}{\bolds\ell}

\newcommand{\dif}{\mathrm{d}}

\newcommand{\mm}{\mathrm{m}}
\newcommand{\xb}{\mathbf{x}}
\newcommand{\rb}{\mathbf{r}}

\makeatother

\begin{document}
\begin{frontmatter}

\title{Gravitational Lensing Accuracy Testing 2010 (GREAT10) Challenge Handbook}
\runtitle{GREAT10 Handbook}

\begin{aug}
\author[A]{\fnms{Thomas}~\snm{Kitching}\corref{}\thanksref{t1}\ead[label=e1]{tdk@roe.ac.uk}},
\author[B]{\fnms{Adam}~\snm{Amara}\thanksref{t8}},
\author[C,P]{\fnms{Mandeep}~\snm{Gill}},
\author[D]{\fnms{Stefan}~\snm{Harmeling}},
\author[A]{\fnms{Catherine}~\snm{Heymans}\thanksref{t2}},
\author[A]{\fnms{Richard}~\snm{Massey}\thanksref{t4}},
\author[E]{\fnms{Barnaby}~\snm{Rowe}\thanksref{t6}},
\author[F]{\fnms{Tim}~\snm{Schrabback}},
\author[G]{\fnms{Lisa}~\snm{Voigt}},
\author[H]{\fnms{Sreekumar}~\snm{Balan}},
\author[I]{\fnms{Gary}~\snm{Bernstein}},
\author[D,J]{\fnms{Matthias}~\snm{Bethge}},
\author[G]{\fnms{Sarah}~\snm{Bridle}\thanksref{t3}},
\author[K]{\fnms{Frederic}~\snm{Courbin}\thanksref{t5}},
\author[K]{\fnms{Marc}~\snm{Gentile}\thanksref{t5}},
\author[A]{\fnms{Alan}~\snm{Heavens}},
\author[D]{\fnms{Michael}~\snm{Hirsch}},
\author[D]{\fnms{Reshad}~\snm{Hosseini}},
\author[A]{\fnms{Alina}~\snm{Kiessling}\thanksref{t6}},
\author[G]{\fnms{Donnacha}~\snm{Kirk}},
\author[F]{\fnms{Konrad}~\snm{Kuijken}},
\author[L]{\fnms{Rachel}~\snm{Mandelbaum}},
\author[E]{\fnms{Baback}~\snm{Moghaddam}},
\author[K]{\fnms{Guldariya}~\snm{Nurbaeva}\thanksref{t5}},
\author[M]{\fnms{Stephane}~\snm{Paulin-Henriksson}},
\author[M]{\fnms{Anais}~\snm{Rassat}},
\author[E]{\fnms{Jason}~\snm{Rhodes}},
\author[D]{\fnms{Bernhard}~\snm{Sch\"olkopf}},
\author[G]{\fnms{John}~\snm{Shawe-Taylor}},
\author[C]{\fnms{Marina}~\snm{Shmakova}},
\author[A]{\fnms{Andy}~\snm{Taylor}},
\author[F]{\fnms{Malin}~\snm{Velander}\thanksref{t6}},
\author[N]{\fnms{Ludovic}~\snm{van Waerbeke}},\\
\author[G]{\fnms{Dugan}~\snm{Witherick}\thanksref{t7}} and
\author[O]{\fnms{David}~\snm{Wittman}}

\runauthor{T. Kitching et al.}

\affiliation{University of Edinburgh,
ETH, Zurich,
Kavli Institute for Particle Astrophysics and Cosmology and Ohio State University,
University T\"{u}bingen,
University of Edinburgh,
University of Edinburgh, California Institute of Technology,
Leiden University,
University College London,
University College London,
University of Pennsylvania,
Max Planck Institute for Biological Cybernetics, University T\"{u}bingen,
    and Institute for Theoretical Physics,
University T\"{u}bingen,
University College London,
Ecole Polytechnique Federale de Lausanne (EPFL),
Ecole Polytechnique Federale de Lausanne (EPFL),
University of Edinburgh,
University T\"{u}bingen,
University T\"{u}bingen,
University of Edinburgh,
University College London,
Leiden University,
Princeton University,
California Institute of Technology,
Ecole Polytechnique Federale de Lausanne (EPFL),
CEA-Saclay, Service d'Astrophysique, Paris,
CEA-Saclay, Service d'Astrophysique, Paris,
California Institute of Technology,
University T\"{u}bingen,
University College London,
Kavli Institute for Particle Astrophysics and Cosmology,
University of Edinburgh,
Leiden University,
Univeristy of British Columbia,
University College London
    and University of California, Davis}

\address[B]{A. Amara\\
ETH, Zurich}
\address[O]{D. Wittman\\
University of California, Davis\hspace*{31pt}}
\address[H]{S. Balan\\
Cavendish Laboratory\\
University of Cambridge}
\address[I]{G. Bernstein\\
Department of Physics and Astronomy\\
University of Pennsylvania}
\address[J]{M. Bethge\\
Institute for Theoretical Physics\\
University T\"{u}bingen}
\address[L]{R. Mandelbaum\\
Department of Astrophysical Sciences\\
Princeton University}
\address[N]{L. van Waerbeke\\
Univeristy of British Columbia\\
Vancouver}
\address[P]{M. Gill\\
Center for Cosmology\\
\quad and AstroParticle Physics\hspace*{30.3pt}\\
Physics Department\\
Ohio State University}
\address[F]{T. Schrabback\\
K. Kuijken\\
M. Velander\\
Leiden Observatory\\
Leiden University}
\address[C]{M. Gill\\
M. Shmakova\\
Kavli Institute\\
\quad for Particle Astrophysics\hspace*{32.7pt}\\
\quad and Cosmology\\
Stanford}
\address[G]{L. Voigt\\
S. Bridle\\
D. Kirk\\
J. Shawe-Taylor\\
D. Witherick\\
University College London}
\address[M]{S. Paulin-Henriksson\\
A. Rassat\\
CEA-Saclay\\
Service d'Astrophysique\hspace*{49.6pt}\\
Paris\\
France}
\address[K]{F. Courbin\\
M. Gentile\\
G. Nurbaeva\\
Laboratoire d'Astrophysique\\
Ecole Polytechnique Federale\\
\quad de Lausanne (EPFL)}
\address[E]{B. Rowe\\
B. Moghaddam\\
J. Rhodes\\
Jet Propulsion Laboratory\\
California Institute of Technology\\
Pasadena, California\\
USA}
\address[D]{S. Harmeling\\
M. Bethge\\
M. Hirsch\\
R. Hosseini\\
B. Sch\"olkopf\\
Max Planck Institute\\
\quad for Biological Cybernetics\\
University T\"{u}bingen}
\address[A]{T. Kitching\\
C. Heymans\\
R. Massey\\
A. Heavens\\
A. Kiessling\\
A. Taylor\\
Institute for Astronomy\hspace*{46.63pt}\\
University of Edinburgh\\
Royal Observatory\\
Blackford Hill\\
Edinburgh, EH9 3HJ\\
United Kingdom\\
\printead{e1}} 
\end{aug}

\thankstext{t1}{Supported by STFC Rolling Grant Number RA0888 and an RAS 2010 Fellowship.}

\thankstext{t8}{Supported by the Zwicky Fellowship at ETH Zurich.}

\thankstext{t2}{Supported by the European Research Council under the EC FP7 ERC Grant Number 240185.}

\thankstext{t4}{Supported by STFC Advanced Fellowship \#PP/E006450/1 and ERC Grant MIRG-CT-208994.}

\thankstext{t6}{Supported by the European DUEL Research-Training Network (MRTN-CT-2006-036133).}

\thankstext{t3}{Supported by a Royal Society University Research Fellowship and a European Research
Council Starting grant.}

\thankstext{t5}{Supported in part by the Swiss National Science Foundation (SNSF).}

\thankstext{t7}{Supported by STFC Grant ST/H008543/1.}

\received{\smonth{9} \syear{2010}}
\revised{\smonth{3} \syear{2011}}

%
\begin{abstract}
GRavitational lEnsing Accuracy Testing 2010 (GREAT10) is\break a public image
analysis challenge aimed at the development of algorithms to analyze
astronomical images. Specifically, the challenge is to measure varying
image distortions in the presence of a variable convolution kernel,
pixelization and noise. This is the second in a series of challenges
set to the astronomy, computer science and statistics communities,
providing a structured environment in which methods can be improved and
tested in preparation for planned astronomical surveys. GREAT10 extends
upon previous work by introducing variable fields into the challenge.
The ``Galaxy Challenge'' involves the precise measurement of galaxy shape
distortions, quantified locally by two parameters called shear, in the
presence of a known convolution kernel. Crucially, the convolution
kernel and the simulated gravitational lensing shape distortion both
now vary as a function of position within the images, as is the case
for real data. In addition, we introduce the ``Star Challenge'' that
concerns the reconstruction of a~variable convolution kernel, similar
to that in a typical astronomical observation. This document details
the GREAT10 Challenge for potential participants. Continually updated
information is also available from
\href{http://www.greatchallenges.info}{www.greatchallenges.info}.
\end{abstract}

%
\begin{keyword}
\kwd{Statistical inference}
\kwd{imaging processing}
\kwd{cosmology}.
\end{keyword}

\end{frontmatter}

\section{Introduction}

The GRavitational lEnsing Accuracy Testing (GREAT) chal\-lenges are a
series of simulations that provide an environment within
which image analysis algorithms of importance for gravitational
lensing cosmology can be developed.
The central theme of GREAT10 is \textit{variability},
the simulations contain spatially variable quantities, and the
challenge is to reconstruct the properties of these variable fields
to a high accuracy.

Gravitational lensing is the effect that light rays are deflected by
gravity. Every galaxy image is distorted by this effect
because of mass that is always present between the galaxy and the observer.
For the majority of galaxies this distortion causes a small additional ellipticity called
shear.
Measuring the shear allows us to extract information on the nature of
the intervening lensing mass and the expansion of the Universe [see
\citet{2001PhR340291B} for a technical review], in particular,
shear can be used to illuminate the nature of dark matter, dark energy
and possible deviations from general relativity.
However, to enable gravitational lensing data
to find this information, the shear needs to be determined to
a high degree of accuracy.

The GREAT challenges are designed to aid in
the development of algorithms, that
aim to measure the gravitational lensing shear, by evolving lensing image
simulations in a controlled manner.
The first of the GREAT challenges, GREAT08
[Bridle et al. (\citeyear{2009AnApS36B})],
began with the zeroth-order problem of measuring a spatially
constant shear in the presence of a spatially constant
convolution kernel (or Point Spread Function, PSF, impulse response).
GREAT08 was inspired by the Shear TEsting Programme [STEP1,
\citet{2006MNRAS3681323H}; STEP2, \citet{2007MNRAS37613M}],
which was a suite of simulations
created by the lensing community and analyzed internally.
GREAT08 was set as a PASCAL\setcounter{footnote}{8}\footnote{\url
{http://www.pascal-network.org/}.}
challenge to both the astronomy and computer science
communities, to encourage the development of interdisciplinary
approaches to this image processing problem, and was
successful on various levels [\citet{2009AnApS36B}]. A high accuracy
was achieved under
the majority of simulated conditions, and there was participation from
outside the gravitational lensing and cosmology communities;
indeed, the winner [Hosseini and Bethge (\citeyear{Hosseini})] was not a
cosmologist.

In GREAT10 both the shear and the PSF are spatially varying
quantities. The shear varies naturally across astronomical images
because of the large scale distribution of matter in the Universe.
The PSF varies spatially in images
because of atmospheric effects and telescope optics.
The primary aim for participants in GREAT10 will be to reconstruct the
correlation function, or power spectrum, of the shear
variation in the presence of a known but varying convolution kernel.

GREAT10 is a PASCAL challenge, set by cosmologists to the
astronomy, computer science and statistics
communities. The challenge will be launched in late 2010 and
will run for 9 months.
Algorithms that are successful when applied to these simulations
will help cosmologists to exploit the scientific potential of current
and future imaging surveys. These surveys will generate petabytes of
imaging information ideal for gravitational lensing, and will
necessitate automated data analysis.

In this article we will introduce the GREAT10 simulations in Section
\ref{The Simulations} and go into some detail in Section
\ref{Simulation Details}.
We will also
outline the submission process, and procedures through which we will
evaluate results in Section~\ref{The Challenge}.
In Section~\ref{GREAT10 Simplifications and Future Challenges} we
conclude by discussing the scope and context of the simulations in
relation to real data. We include a number of \hyperref[app]{Appendices}
that contain
technical details. We summarize the simulation and challenge details
of GREAT10 in Figure \ref{summer}.

\section{The simulations}
\label{The Simulations}

\begin{figure}

\includegraphics{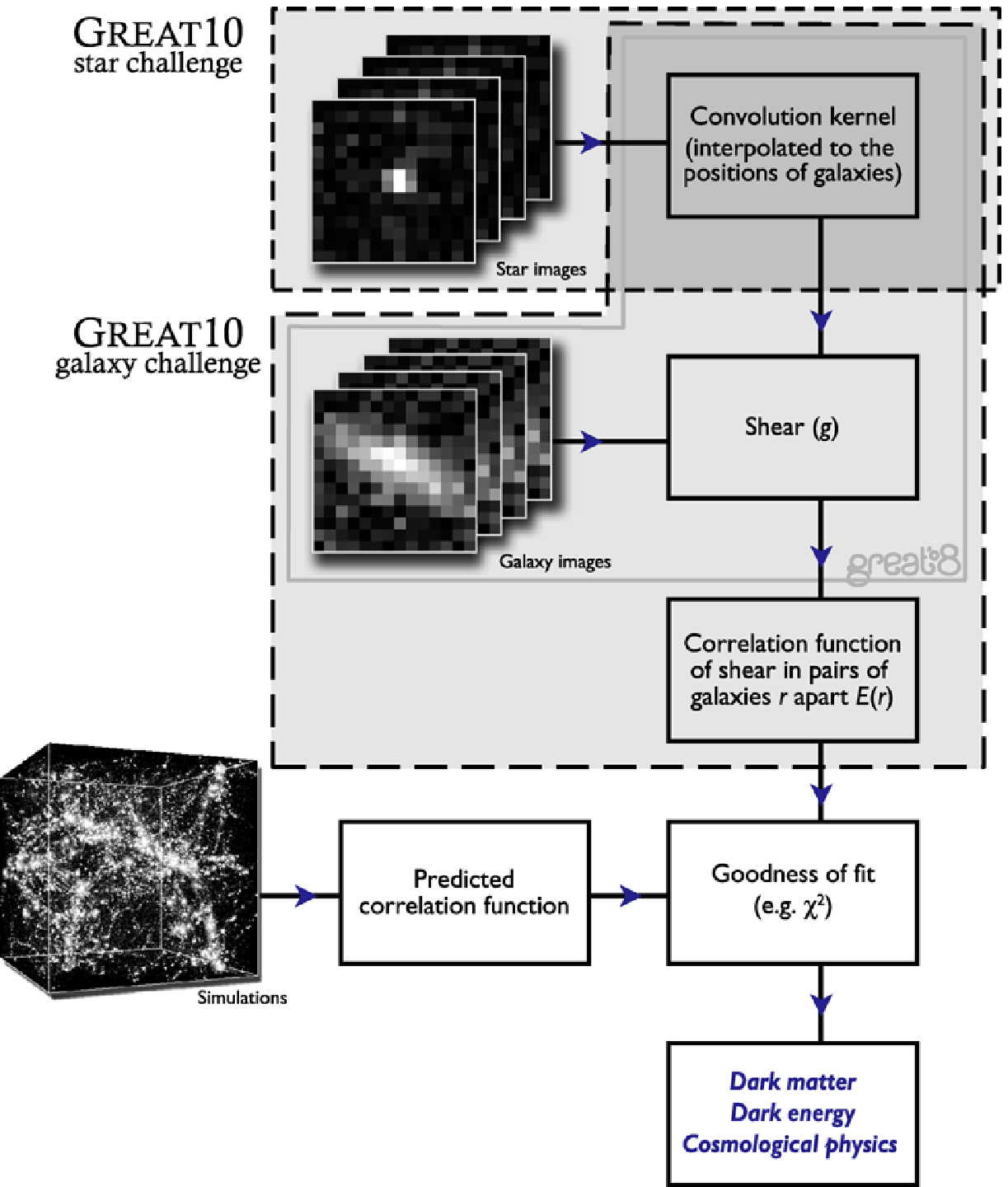}

\caption{GREAT\textit{10} introduces two parallel challenges, both
based on
steps in the analysis of real astronomical data.
The ``Star Challenge'' concerns the shapes of scattered stars. Each star
is a local, noisy realization of the convolution kernel that is
inevitably applied to any astronomical image (due to a combination of
the Earth's atmosphere, telescope optics and detector
imperfections). This kernel varies smoothly across an image, and the
challenge is to interpolate measurements between the locations of
stars.
The ``Galaxy Challenge'' concerns galaxy shapes, which must
be measured after effectively deconvolving the image from a
(supplied) model of the convolution kernel. This builds most closely
on the previous GREAT\textit{08} challenge, but includes the crucial
development that both the kernel and the distortion applied to the
galaxy shapes now vary across an image. The required output is
a~measurement of the correlation between the shapes of pairs of
galaxies separated by various distances.
In a real cosmological analysis, the two procedures are applied in
turn, then the correlations are compared to predictions of detailed
simulations of the Universe.}
\label{Flow}
\end{figure}

In GREAT10 we will introduce \textit{variable fields} into the
shear estimation challenge. Both the shear and the PSF
will vary across the images.
In Figure \ref{Flow} we show how GREAT10 is related to the
process through which we go from
images of galaxies and stars to cosmological parameter
estimation. There are a number of steps in this process that we will
discuss in detail in this section.

%
%
\begin{figure}

\includegraphics{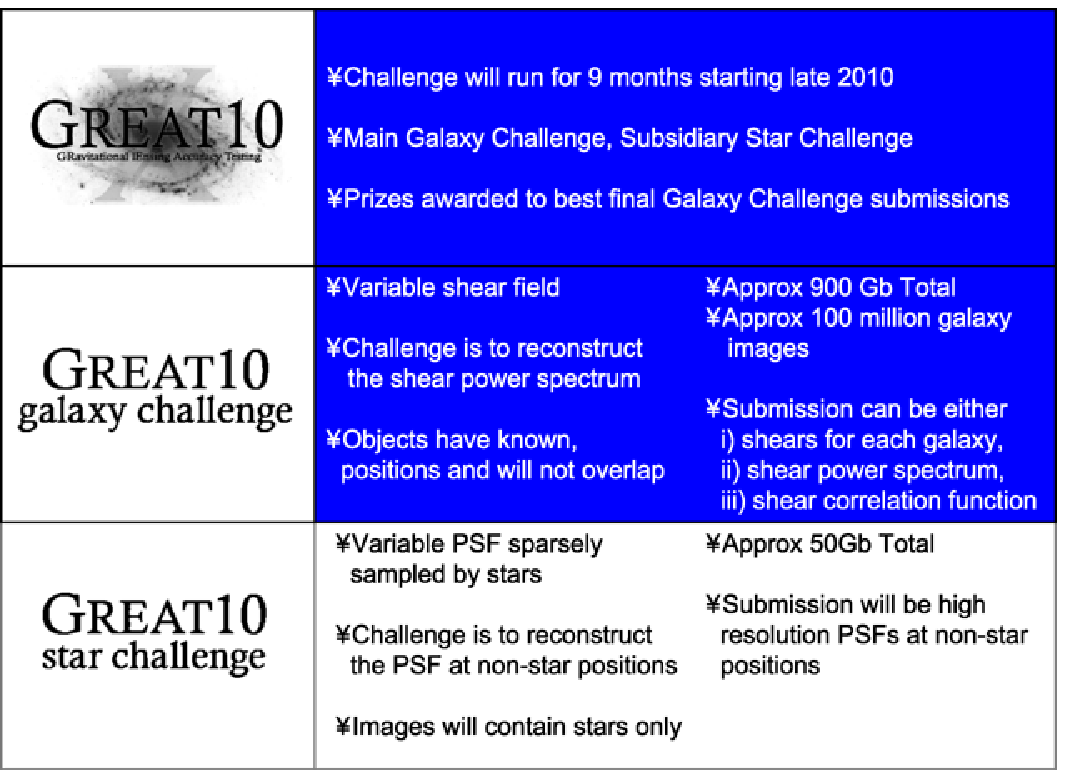}

\caption{Top: A summary of the simulation and challenge details of
GREAT\textit{10}, which will consist of two separate but interconnected
challenges, the main ``Galaxy Challenge'' and the ``Star
Challenge.''}
\label{summer}
\end{figure}

%
%

To preface this section,
we summarize the GREAT10 image simulation steps that represent
each step of the lensing process:
\begin{itemize}
\item
Undistorted Image: We start with an unsheared set of galaxies (with a
distribution of ellipticities described in Appendix \ref{appA}) or
point sources (stars).
\item
Shear Field Applied: The galaxy images, but not the star images,
are transformed using a local
distortion matrix that varies spatially across the image, a ``shear''
field.
\item
Convolution: Both the sheared galaxy images and star images are
locally convolved by a Point Spread Function that varies spatially
across the image.
\item
Pixelization: The convolved star and galaxy images are pixelized by
summing the
intensity in square pixels.
\item
Noise: Uncorrelated Gaussian distributed (homoskedastic) noise
is added to all images. The image simulation process will also create
Poisson noise.
\end{itemize}
Further details are given in Figures \ref{summer}, \ref{Flow} and \ref{forward}.
In this section we will describe the cosmologically important shear
field and the type of convolution kernel that images experience.

\subsection{Gravitational shear}

As a photon propagates from a galaxy to an observer, the path that the
light takes is distorted, or lensed, by the presence of mass along the
line of sight. The first-order effect that this gravitational lensing has
on the image of a galaxy
is to introduce a local distortion that can be expressed as a
remapping of the unlensed, original, pixels
%
%
\begin{equation}
\label{eq1}
\pmatrix{x_{\mathrm{original}}\cr
y_{\mathrm{original}}}
=
\left[\pmatrix{1&0\cr
0&1}
-|g|
\pmatrix{
\cos(2\phi)&\sin(2\phi)\cr
\sin(2\phi)&-\cos(2\phi)}\right]
\pmatrix{x_{\mathrm{observed}}\cr
y_{\mathrm{observed}}},
\end{equation}
where $x$ and $y$ denote a coordinate system that describes the
observed light distribution of an object.\footnote{We highlight a caveat
that, in
general, gravitational lensing does also introduce a term that alters
the observed size of an object. This term modifies
equation \eqref{eq1} by introducing a multiplication factor $(1-\kappa)$
where $\kappa$ is called the ``convergence.'' In GREAT10 we explicitly
set $\kappa=0$ in all images.} The second term in this expression
is known as the \textit{shear}, and has the effect of stretching an object
in a direction parameterized by an angle $\phi$ by an amount $|g|$.

It is important to note that the factor of $2$ in angle means that the
shear is rotationally symmetric under rotations of $180$ degrees.
In general, we define the shear as a local complex variable
%
%
\begin{equation}
\label{eq11}
g=|g|\mathrm{e}^{2\mathrm{i}\phi}=|g|[\cos(2\phi)+\sin(2\phi
)]=g_1+\mathrm{i}g_2,
\end{equation}
where $g_1$ represents
local ``$+$ type'' distortions (along the Cartesian axes) and $g_2$
represents local
``$x$ type'' distortions (along the $45$ degree axes). Equation \eqref
{eq1} can
now be rewritten in terms of $g_1$ and $g_2$ as
%
%
\begin{equation}
\label{eq12}
\pmatrix{
x_{\mathrm{original}}\cr
y_{\mathrm{original}}}
=
\pmatrix{
1-g_1&-g_2\cr
-g_2&1+g_1}
\pmatrix{
x_{\mathrm{observed}}\cr
y_{\mathrm{observed}}}.
\end{equation}
The amplitude and direction of the shear $g$ that an object
experiences depends on the amount and nature of the lensing matter,
rendering this quantity of great interest to the cosmological
community.

The shear effect, to first order,
produces an additional \textit{ellipticity} in an
object image; here we define the
ellipticity of an object as $e=(1-r)(1+r)^{-1}\exp(2\mathrm{i}\psi)$,
where $r$ is the ratio of the major to minor axes in the image, and
$\psi$ is an angle of orientation.
A significant characteristic of gravitational lensing is
that galaxies in general, and in
GREAT10, already have an ``intrinsic'' presheared ellipticity $ e^{
\mathrm{intrinsic}}$ such that the
measured ellipticity per galaxy can be expressed as\footnote{Here we
use $|e|=(1-r)(1+r)^{-1}$ for the ellipticity, alternatively,
one can use $|e|=(1-r^2)(1+r^2)^{-1}$ which leads to an extra factor
of $2$ in \eqref{eint}; see \citet{2001PhR340291B},
Section 4.2, for a detailed discussion.}
%
%
\begin{equation}
\label{eint}
e^{\mathrm{observed}}=\frac{e^{\mathrm{intrinsic}}+g}{1+g^*e^{
\mathrm{intrinsic}}}\approx e^{\mathrm{intrinsic}}+g,
\end{equation}
where the approximation is true for small shear $|g|\ll1$; in GREAT08
and GREAT10 the shear has values of $|g|\ls0.05$.
One aspect of the
challenge for gravitational lensing is that we cannot directly observe
the unsheared images of the galaxies that we use in the analyses.

\subsection{Variable shear fields}

The shear we observe varies spatially as a~function of
position on the sky. This
variation reflects the distribution of mass in the
Universe, which forms a ``cosmic web'' of structures
within which galaxies cluster on all scales through the influence of
gravity. As we observe galaxies, at different positions and at
different distances, through this cosmic web the light from each
galaxy is deflected by a~different amount by a different distribution of
mass along the line of sight. The effect is that the shear induced on
galaxy images is not constant across the sky but instead varies in a
way that reflects the large scale, nonuniform, distribution of mass
in the Universe. This lensing by large scale structure is
known as \textit{cosmic shear}. The measurement of cosmic shear has been
one of the major goals of gravitational lensing and promises to become
one of the
most sensitive methods with which we will learn about dark energy and
dark matter.

The shear field we observe
is in fact a \textit{spin-2} field. This means that it is a~scalar
field described locally by $g_1$ and
$g_2$ components that introduce $180$~de\-gree invariant distortions [the
``$2$'' in ``spin-2'' refers to the factor of 2 in (\ref{eq1})
and (\ref{eq11})].
In this case\vspace*{1pt} equation ({\ref{eq11}) acts as a local
coordinate transformation for each galaxy where $g_1(\theta_x,\theta
_y)$ and
$g_2(\theta_x,\theta_y)$ are now variable
fields as a~function of position\footnote{In GREAT10 we will assume
that the shear is constant
on scales of the galaxy images themselves, such that the lensing effect
is a local
coordinate transform as in (\ref{eq1}). However, we note that
a second-order
weak lensing effect called \textit{flexion} [\citet{2005ApJ619741G}]
is present in real data and in very
high mass regions the lensing can even produce arcs and multiply imaged galaxies
(called strong lensing).} \mbox{($\theta_x$, $\theta_y$)}.

A general spin-2 field can be written as a sum of an \textit{E-mode}
component (curl-free) or gradient term, and a \textit{B-mode} component
(grad-free) or curl term; E and B are used in analogy to the electric
and magnetic
components of an electromagnetic field. The cosmological effect on the
shear field is to introduce an E-mode
signal and only negligible B-mode, which means that a tangential shear
is induced around any
region of excess density (see Figure~\ref{varshear}).\looseness=-1

In GREAT10 we will evaluate methods on their ability to reconstruct
the \mbox{E-mode} variation only.
In real observations the unsheared orientations of galaxies
will in general produce an E- and B-mode in the variation of the
observed ellipticities [equation (\ref{eint})].
However, in GREAT10 the unsheared population of
galaxies will have a pure B-mode ellipticity distribution. This B-mode
is introduced to reduce ``ellipticity noise terms'' in the simulations,
and means that the size of the simulations is significantly reduced;
this B-mode will not contribute to the E-mode variation with which we
will evaluate the results. The intrinsic
(undistorted) shape of a galaxy is not observable, galaxies always
have the additional lensing distortion. The uncertainty on our
knowledge of the intrinsic ellipticity of a galaxy, which is commonly
related to the intrinsic variance in the ellipticities of the population,
is what is refered to as ``ellipticity noise.''
We present a technical discussion of this aspect of the simulations in
Appendix \ref{appA}.

\subsection{Measuring a variable shear field}

The galaxies we observe act as discrete points at which the shear field
is sampled.
The amount of shear induced on the observed image of an individual
object is
small, typically $\ls$3\% (a change in the major-to-minor axis ratio
of $\sim$1.06)
and so techniques have been developed
to measure statistical features of the shear distribution from an
ensemble of galaxies.
The mean of the shear field $\langle g_i\rangle$ is zero when averaged over
sufficiently large scales,
however, the variance of the shear field is nonzero and
contains cosmological information.
We wish to calculate the variance of the
shear field as a function of scale (small scale means in close
proximity in an image);
a large variance on small scales, for
example, would signify a matter distribution with structures on those
scales (see
Figure \ref{realobs}).

%
%
\begin{figure}

\includegraphics{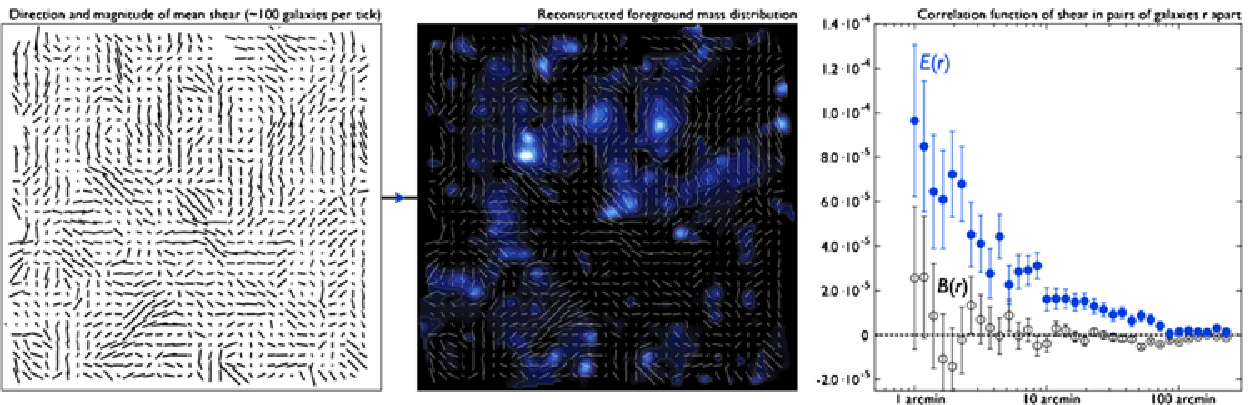}

\caption{\textup{Real lensing effect on an ensemble of
galaxy images}. The shear distortion $g=g_1+\mathrm{i}g_2$ measured
from the shapes of galaxies is represented in the left panel as a~tick
mark in the direction of $g$, with a length proportional to its
magnitude. The amount of shear in a particular region of the image is
determined by the amount of lensing material (dark matter) along the
line of sight from the galaxy. Hence, we can reconstruct a~map of the
mass shown in the middle panel. We can also take the two-point
correlation function of the shear field which is shown in the right
panel. The left and middle panels show data from the Hubble COSMOS
survey [Massey et al. (\protect\citeyear{2007MNRAS37613M})] and the right panel
shows a~correlation function from the CFHT Legacy Survey
[Fu et al. (\protect\citeyear{2008AA4799F})].} \label{realobs}
\end{figure}

%
%
\begin{figure}

\includegraphics{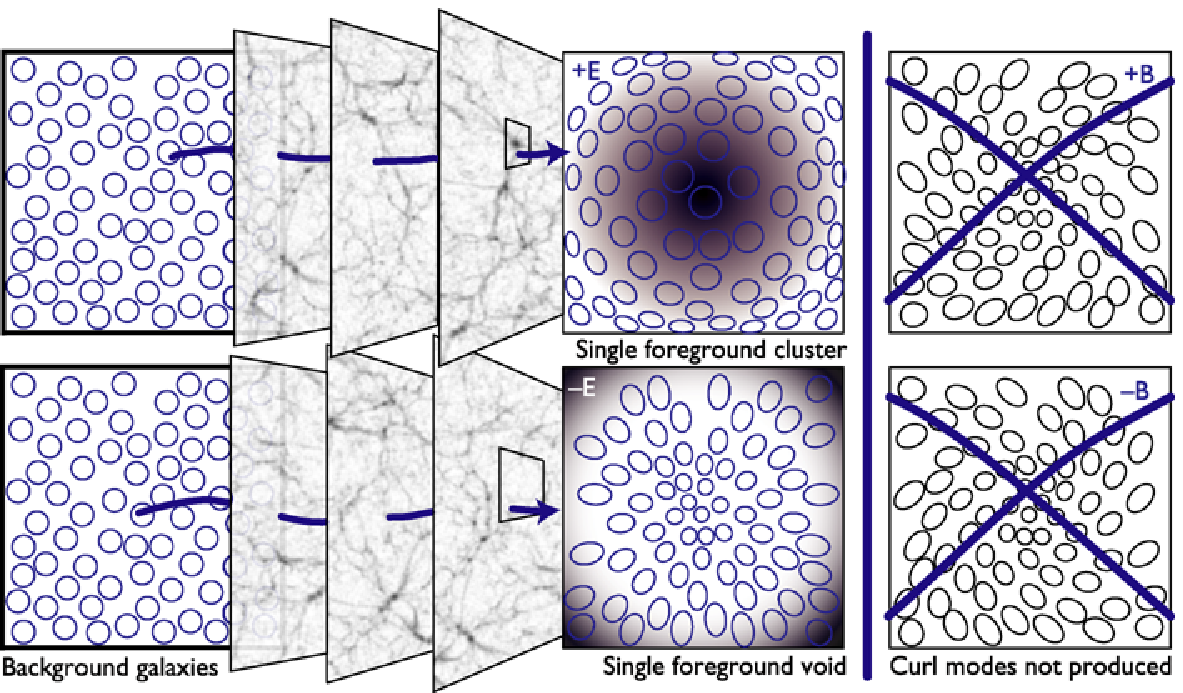}

\caption{\textup{Schematic effect on an ensemble of galaxy images}.
The shear distortion varies as a
function of position because the light propagates
through the variable, foreground large scale structure (LSS) or
``cosmic web.''
The properties of the
LSS imprint specific signatures on the type of distortion pattern.
For example, if we had a sample of circular background galaxies
(clear circles, left
panels), they would be distorted into ellipses with major axes
orientated tangent to density contours by a
foreground mass overdensity, or radial shapes by a mass
underdensity. These typically contribute to the cosmic shear on different
physical scales, and their combined effect is an ``E-mode'' correlation
function that varies as a function of scale. However,
the LSS does not introduce significant ``B-mode'' or
curl distortion patterns (represented by a cross X in this
figure).}
\label{varshear}
\end{figure}

We can calculate the variance of the shear field as a function of
scale by computing the
two-point \textit{correlation function} $\xi(\Delta\thetab)$ of the
shear field.
The correlation function measures the tendency for galaxies at a
chosen separation to have preferred shape alignment.
For any given pair of galaxies we can define a component of the shear
from each in terms of a tangential shear component,
which is perpendicular to the line joining the pair, and a~cross-component which is at $45$
degrees, this is meant to isolate the E- and B-mode signal shown in
Figure \ref{varshear}. In Appendix~\ref{appB} we show how to calculate the
correlation function from galaxy shears.

Complementary to the correlation function is the \textit{power spectrum}
$C(\bell)$, which is simply the
Fourier transform of the correlation function,
%
%
\begin{eqnarray}
\xi(\Delta\thetab)&=&\int C(\bell)\mathrm{e}^{\mathrm{i}\bell.\Delta
\thetab}
\,\mathrm{d}^2\bell,\nonumber\\[-8pt]\\[-8pt]
C(\bell)&=&\frac{1}{(2\pi)^2}\int\xi(\Delta\thetab){\mathrm
e}^{-{\mathrm i}\bell.\Delta\thetab}
\,\mathrm{d}^2\Delta\thetab,\nonumber
\end{eqnarray}
where $\bell=(\ell_x$, $\ell_y)$ is a 2D wavevector probing scales
of order $\Delta\thetab=(2\pi)/|\bell|$.
The shear power spectrum can again be related to the underlying matter
distribution. In Appendix \ref{appB} we outline a simple method for creating
the shear power spectrum from the local $g_1$ and $g_2$ shear
estimators from each galaxy. When the simulations are made public we
will also provide open-source code that will enable participants
to create any of these statistics from a catalogue of discretely
estimated shear values.

The reconstruction of the shear field variation to sufficient
accuracy, in terms of the power
spectrum or correlation function, is the GREAT10 \textit{Galaxy Challenge}.

\subsection{Simulating a variable shear field}

In GREAT10 we will simulate the shear field as a \textit{Gaussian random
field}. These fields have a random distribution of phases,
but a distribution of amplitudes described by the input power
spectrum. These are simplified simulations relative to the real shear
distribution. In particular, the filamentary structures seen in the
``cosmic web'' are
due to more realistic effects, and will not be present in the GREAT10
simulations. In Figure \ref{simshear} we show an example of the
simulated shear field in a GREAT10 image.

%
%
\begin{figure}

\includegraphics{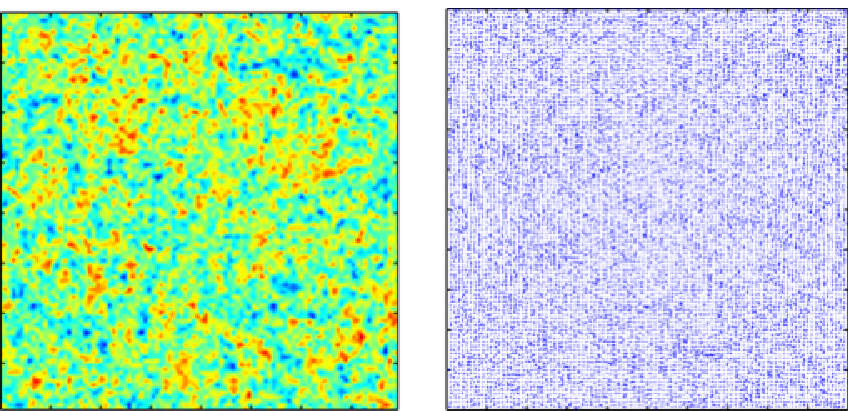}

\caption{\textup{A GREAT10 simulated shear variation}. An example of the
variable shear field in a~GREAT\textit{10} image. Left panel: we show the
simulated mass distribution, which is a~Gaussian realization with a
particular power spectrum. Right panel: we show the shear
represented by a line with length proportional to the amplitude of
the shear $|g|$ and an angle $\phi$; see \protect\eqref{eq1} and
\protect\eqref{eq11}.}
\label{simshear}
\end{figure}

In addition to the cosmologically interesting shear
effect, galaxy images are smoothed by a spatially
variable convolution kernel (PSF).

\subsection{Variable PSF}
\label{Variable PSF}

The images that we use in gravitational lensing analyses are smoothed
and distorted by a convolution kernel (or Point Spread Function, PSF).
The PSF that these images are convolved with is produced by a
combination of effects:
\begin{itemize}
\item
The images we use are created by observations with telescopes, which
have a characteristic PSF that can vary as a function of position in the
image due to the exact optical setup of the telescope and camera. Also, the
detectors we use, Charge Coupled Devices (CCDs), pixelize the image;
and defects can cause image degradation.
\item
When making observations with the telescopes on the ground the
atmosphere acts to induce an additional PSF (due to refraction in the
atmosphere and turbulence along the path of the photons).
\item
A telescope may move slightly during an observation, adding an additional
smoothing component to the PSF. Typical
observations last for seconds or minutes during which various
effects (such as wind, temperature gradients, vibrations, etc.) can cause
the telescope to move.
\end{itemize}
Each of these effects can vary across an image in either a
deterministic fashion (as in telescope optics) or randomly (as in the
atmosphere).
In Figure~\ref{PSFrealobs} we show the effect on an
ensemble of stars, both schematically and from realistic
simulations. In Figure \ref{simpsf} we show the spatial variation of a
simulated GREAT10 PSF.

%
%
\begin{figure}

\includegraphics{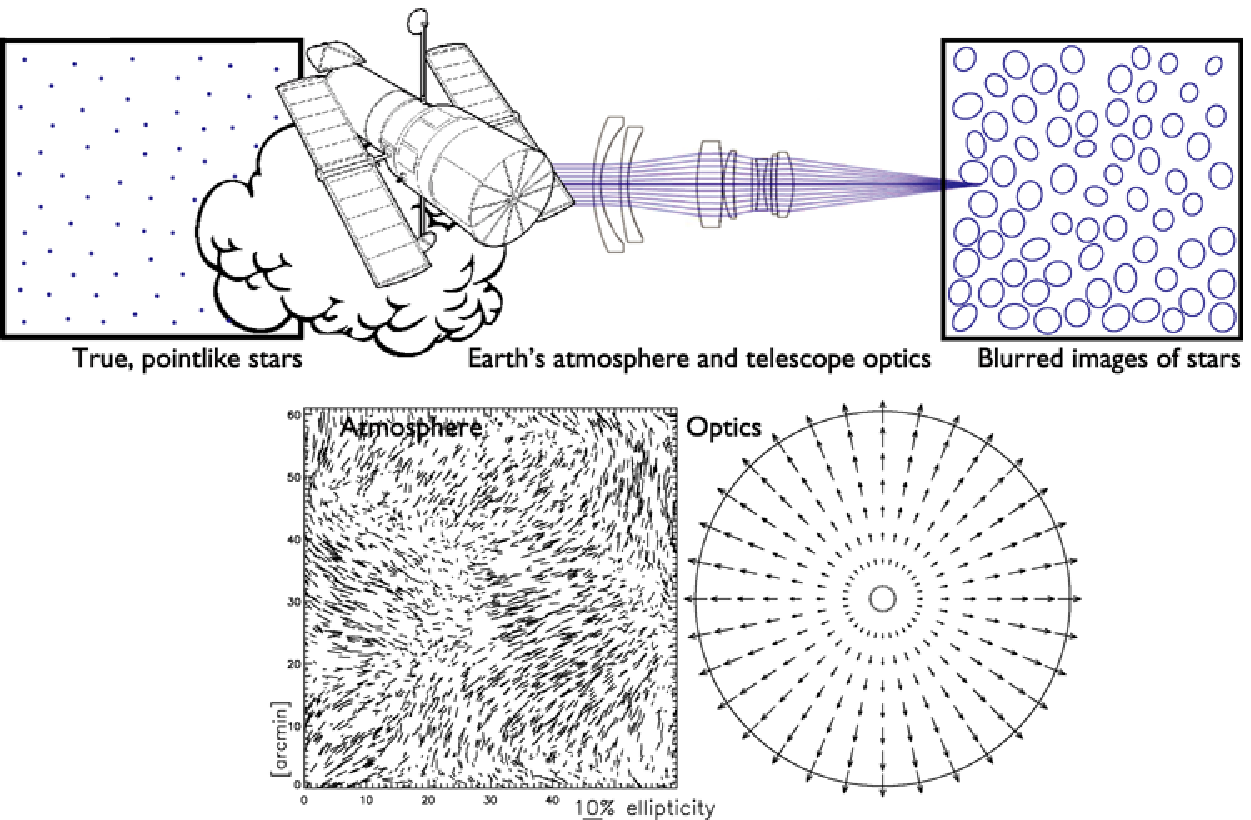}

\caption{Upper panels: \textup{Schematic Effect} of
atmosphere and telescope optics. A distribution of point-like star
images is blurred and the resulting pattern has variable spatial
structure. Lower panels: \textup{Real Effect} of the atmosphere [from
Rowe (\protect\citeyear{2010MNRAS404350R})] and telescope optics [example from
Jarvis, Schechter and Jain (\protect\citeyear{2008arXiv08100027J})] on the PSF
ellipticity and size, respectively. The atmosphere adds random,
coherent, patterns and the telescope adds specific functional behavior
due to optical effects.} \label{PSFrealobs}
\vspace*{4pt}
\end{figure}

We can estimate
the local PSF, in the presence of pixelization, from images of stars. This
is because stars are point-like objects that only experience the
convolving effect of the PSF and are not subject to the shear
distortion; the stars we observe are part of our own galaxy, the Milky
Way. The problem is that we have a discrete number of point-like
PSF estimators in each image and a spatially varying PSF. What we need
is an
accurate PSF reconstruction at nonstar positions (the positions of
the sheared galaxies).

There are currently two broad classes of approach that we define
here that have been investigated to deal with this sparsely sample
variable PSF:
\begin{itemize}
\item
Direct modeling uses star images as discrete estimators of a
spatially varying model (typically some low-order polynomial) and finds a
best-fit solution to this model.
%
\item
Indirect modeling, for example, uses an ensemble of stellar images
at different positions to extract principal components (or
eigenvectors) of the PSF across the images.
\end{itemize}
These techniques, direct and indirect modeling,
have been used on gravitational lensing
images, but none have yet reached the required accuracy to fully
exploit the most ambitious future gravitational lensing experiments.

In addition, there are a variety of
deconvolution algorithims that attempt to remove the PSF from images by
applying an
algorithm that reverses the effect of the convolution.
The deconvolution approach has been investigated, but
has not been implemented with gravitational lensing observations to
date. In Appendix \ref{appE} we outline in more detail these three PSF modeling
approaches.

To first order the PSF is commonly parameterized with a size (usually
defined using a Full-Width-Half-Maximum, FWHM, Gaussian measure) and an
ellipticity, defined in a similar way to shear
%
%
\begin{equation}
\label{peq1}
e=|e|[\cos(2\psi)+\sin(2\psi)]=e_1+{\mathrm i}e_2,
\end{equation}
where again $e_1$ represents ``$+$ type'' distortions and $e_2$ represents
``$x$ type'' distortions.
In the GREAT10 Galaxy Challenge the PSF will be provided as a known
function $f(e,R)$ parameterized with a size $R(\theta_x,\theta_y)$
(FWHM) and
ellipticity $e(\theta_x,\theta_y)$ that will vary across each image,
so that any inaccuracies caused by PSF misestimation should be removed
from the problem.

%
%
\begin{figure}

\includegraphics{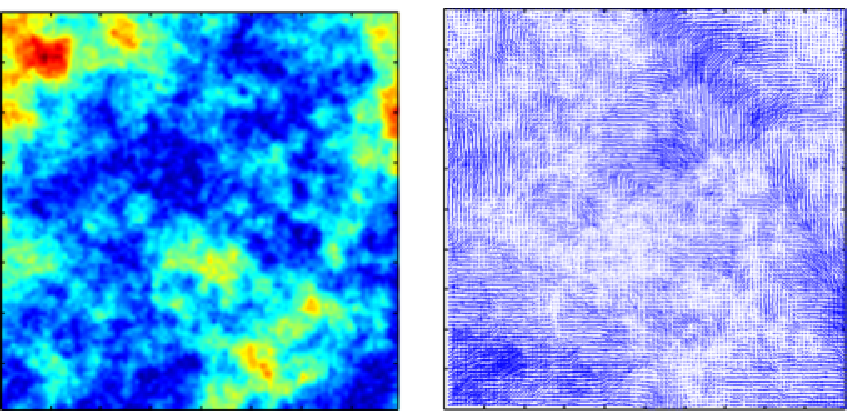}

\caption{\textup{A GREAT10 simulated PSF variation}. An example of the
variable PSF field in a GREAT\textit{10} image. Left panel: we show the
simulated magnitude of the ellipticity, which contains atmospheric
and telescope optics effects. Right panel: we show the ellipcity
represented by a line with length proportional to the amplitude of
the ellipticity $|e|$ and an angle $\psi$; see \protect\eqref{peq1}.}
\label{simpsf}
\end{figure}

However, in addition to measuring shear very accurately,
PSF estimation is of crucial importance for gravitational lensing. If
we cannot characterize the PSF sufficiently at galaxy
positions, our shear values will be inaccurate. This is not
addressed in the Galaxy Challenge where the PSF is a known function,
so here we set the additional task of estimating the PSF at nonstar
positions as a challenge in itself; the GREAT10 \textit{Star Challenge}.

\subsection{Summary of effects}

To summarize the effects included in the simulations,
we show the effect induced on an individual galaxy and star
image in Figure \ref{forward}. This ``forward process,'' from
a galaxy to an image, was
detailed in Bridle et al. (\citeyear{2009AnApS36B}).
Each galaxy image is distorted by the matter distribution, this image
is then convolved by a PSF that spatially varies as a result of
possible atmospheric effects and telescope optics, and finally the image
is pixelized. Star images experience the convolution and pixelization
but are not distorted by the shear field.

%
%
\begin{figure}

\includegraphics{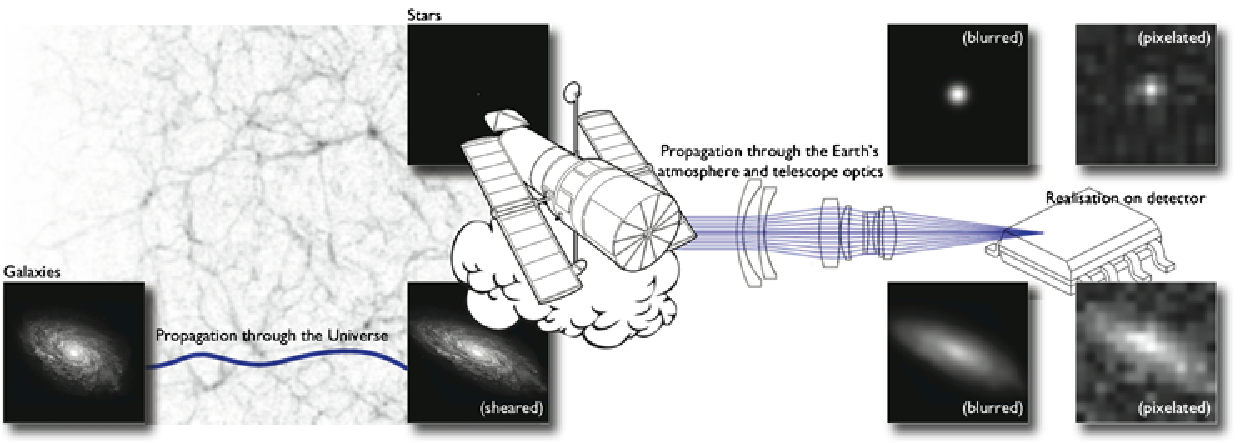}

\caption{\textup{Summary of the main effects on an individual galaxy or star}.
This is the \textup{forward process} described in Bridle et al.
(\protect\citeyear{2009AnApS36B}), although
GREAT\textit{10} includes additional realism in that both the induced
gravitational shear distortion and the PSF vary as a function of position.
The GREAT\textit{10} ``Galaxy Challenge'' is to estimate the shear distortion
applied to a~galaxy image, correcting for the additional effects in the
Earth's atmosphere and telescope optics, which are also experienced
by the images of stars; for space-based telescopes the atmospheric
effects are not present so that telescope and detector
effects alone induce a~PSF.}
\label{forward}
\end{figure}

The GREAT08 challenge focused on the determination of the shear $g$,
in the presence of these effects,
by creating images in which a constant shear and a constant PSF had been
applied to all objects (by creating constant-shear images,
algorithms designed to estimate the average shear from an ensemble of galaxies
can be tested). In GREAT10 we will move toward the more demanding, and
more realistic,
regime of variable shear fields.

In gravitational lensing analyses to date (c. 2010),
we estimate the local shear from
individual galaxies and use these to reconstruct the two-point
statistics of the shear field. In Appendix \ref{appD} we review current
methods that are used to estimate the shear from galaxies [including
some advancements made since GREAT08; \citet{2009AnApS36B}].

\section{Simulation details}
\label{Simulation Details}

In GREAT10 the main Galaxy Challenge is to reconstruct the shear field
in the
presence of a variable PSF. We also present the Star Challenge that is
independent from the main challenge:
\begin{itemize}
\item
Galaxy Challenge. This is the main challenge for GREAT10. The objective is
to reconstruct the shear
power spectrum. This is most similar to GREAT08. In this challenge the
varying PSF will be a known function.
\item
Star Challenge. This separate challenge is to reconstruct the PSF
at nonstar positions given an image of PSF convolved stars.
\end{itemize}
The GREAT10 structure is schematically represented in Figure \ref{summer}.

\subsection{Galaxy challenge}

In each Galaxy Challenge image there will be a~different realization of
the shear power spectrum. The Galaxy Challenge will be
subdivided into a low noise (high signal-to-noise)
set and a realistic (real) noise set. The low noise
subset will be closely matched in substructure to the real noise set
such that participants will be able to analyze
whether training on such data is of use in shape measurement.
There will be fewer low noise images than
real noise images (similar to GREAT08) to reflect our requirements on
accuracy (see Section \ref{Shear Quality Factor}).

For the main Galaxy Challenge the input PSF will be provided
as a functional form that will allow the spatially varying PSF to be
reconstructed at any spatial position with sub-pixel resolution.
Images will be supplied
containing sheared galaxy images, and the approximate positions of each object
will be provided.
The participants will be asked to provide a reconstruction of the shear power
spectrum at specified $\ell$ mode ($2\pi/\mbox{separation}$) values.
Participants will submit either
of the following:
\begin{itemize}
\item
A ``shear catalogue'': a value of $g_1$ and $g_2$ for every
galaxy in each image.\footnote{Nonzero weights for individual objects will
also be allowed.}
\item
A correlation function as a function of separation: at the launch of
the challenge we will specify the exact values and binning required for
this type of submission.
\item
A shear power spectrum as a function of $\ell=2\pi/ \mbox{separation}$:
at the
launch of
the challenge we will specify the exact values and binning required for
this type of submission.
\end{itemize}
%
In addition to the innovations of variable fields,
we will also make the Galaxy Challenge simulation more realistic in that
the distribution of galaxy properties, for example, size and
signal-to-noise, will be continuous distributions rather than discrete
in some
images. The GREAT10 galaxy models are similar to those in GREAT08;
they will consist of two components, a~``bulge'' and
a ``disk'' (an exponential profile with ${\mathrm e}^{-kR^{1/n}}$ with $n=1$
and $n=4$, resp.); these components may be misaligned and
have varying intensities.\looseness=1

\subsection{Star challenge}

In each Star Challenge image there will be a different PSF.
These Star Challenge images will be grouped into sets that
contain the same underlying PSF,
except for the presence of a possible random component that will have
a different realization in each image,
to represent a~series of gravitational lensing observations.
The Star Challenge will only have one high signal-to-noise
level. In real CCDs some bright stars can ``saturate'' the images and
need to be
masked, in GREAT10 there will be no saturated stars.

For the Star Challenge, participants will be asked to provide an
estimation of the PSF at nonstar positions.

All
additional necessary information on the simulations will be provided on
the GREAT10 webpage at the time of launch.
Participants will also be
provided with training sets for each Challenge that will be
exactly representative of some subset of the main Challenge---in real
data we do not have training sets and would have to rely on
simulations being accurate enough representations of the Universe,
however, we invite participants to use the training data with the
caveat that this may only guarantee unbiased results in some subset
of the main Challenge.

\section{The challenge}
\label{The Challenge}

In this section we will summarize some of the practical details of the
Challenge and define our evaluation procedure.

\subsection{Challenge details}

The GREAT10 competition will run for $9$ months.
The competition will be to achieve the largest quality factors
in either the Galaxy and Star Challenges. Each challenge will be run as
a separate
competition. We will award prizes for the largest average quality
factor in the Galaxy Challenge (there
will only be one prize for all types of submission: catalogue, power spectrum,
correlation function). We will also award a prize, in either the Star
or Galaxy Challenge, for a method that
performs well under a variety of simulated conditions or whose
innovation is particularly noteworthy.
There will be a mid-challenge workshop and a final GREAT10 meeting,
where we will present the Challenge prizes.

Participants will be required to download the simulation
data. We will
provide download nodes, hosted by GREAT10 team members,
over various continents for ease of accessibility. The simulated data
for GREAT10 will constitute approximately $900+50$ GB over the Galaxy
and Star
Challenges, respectively.

The submission process for the competition will be through a web
interface similar to GREAT08 [Bridle et al. (\citeyear{2009AnApS36B});
GREAT08 Handbook],
with a live leaderboard of average
quality factors continuously updated. We will also publish a detailed
results paper where the performance of methods will be shown as a
function of various properties of the simulated images.
We outline some rules of the competition in Appendix \ref{appF}.

\subsection{Challenge evaluation}
\label{Shear Quality Factor}
Each submission will result in a shear power spectrum $C(\bell)$
being calculated, either directly by the participant or internally
after a
shear catalogue or correlation function submission.
The submitted power spectrum will then be
compared to the true input power spectrum for each image in the
simulation and a quality factor calculated.

The quality factor for an individual image is defined using the
difference between the submitted power spectrum and the input power spectrum
[\citet{2008MNRAS391228A}], which is related to variance of the
measured and true shears. For image $i$ this is defined as
%
%
\begin{equation}
\label{sigma}
\sigma_i^2=\frac{1}{2\pi}\int_{\ell_{\min}}^{\ell_{\max}} \ell
(\ell+1)|C^{\mathrm{input}}(\ell)-C^{\mathrm{submitted}}(\ell)|\,\mathrm
{d}\ln\ell.
\end{equation}
This is a quantity that has been used in cosmic shear analyses to gauge the
impact of systematic effects on cosmological parameter
constraints. The fact $\ell(\ell+1)$ comes from summing over density
of states in Fourier space.
For GREAT10 we define the Galaxy Challenge quality factor as
%
%
\begin{equation}
\label{QQ}
Q_{\mathrm{GREAT10}}\equiv\frac{\mathcal{N}}{\langle\sigma^2\rangle},
\end{equation}
where the angular brackets denote an average over all images in the
simulation. The numerator $\mathcal{N}$ will be determined subject to the
exact range of scales $[{\ell_{\min}},{\ell_{\max}}]$ in
\eqref{sigma}, that will be defined at the challenge launch.
\textit{The winner of the Galaxy Challenge will be the method that
results in
the \textit{largest average quality} factor over all images}.
In the final analysis of the results we will also consider alternative
measures such as the L-1 and L-2 norm of the shear estimates versus
the true shears. The results of the Star Challenge will also be gauged
by a quality
factor, which we outline in Appendix~\ref{appE}.\looseness=1

The GREAT10 quality factor is different from the GREAT08 factor
[Bridle et al. (\citeyear{2009AnApS36B})], which
used the root-mean-square of the
shear residuals as the denominator.
The GREAT08 quality factor, and goal, was designed primarily with an
additive bias
$\hat g=g+b$ in mind, as discussed in \citet{2009AnApS36B}, $b$ in
this case is a bias that is constant with respect to
$g$. However, both STEP and GREAT08 have shown that multiplicative
correction
$\hat g=g\times f(g)$ is also important, where $f(g)$ can be some
function of $g$ (possibly constant). The GREAT10 quality
factor is sensitive to both these effects, as well as any
misestimation of shear variance contributions ($\sigma^2_{g}$).

\vspace*{5pt}
\section{GREAT10 simplifications and future challenges}
\label{GREAT10 Simplifications and Future Challenges}

$\!\!\!$We expect GREAT10 to have a significant impact on the gravitational lensing
and cosmological communities, enabling the exploitation of the next
generation of experiments. The simulations we have designed present a
unique challenge to computer science, statistics and
astronomy communities;
we require extreme accuracy from a very large data set, and have
limited training data.

In GREAT10 we have set the challenge of estimating the variation of
the gravitational lensing shear signal in the presence of a realistic
PSF model. By advancing the GREAT challenges in this direction, we have
addressed some of the simplified assumptions that were made in
GREAT08. As a result, GREAT10 is a demanding challenge---however, this
is only the second stage in a series of challenges that will work
toward creating realistic gravitational lensing
simulations. Some of the simplified
assumptions in GREAT10 simulations compared to real data include the following:
\begin{itemize}
\item
Gaussian Random Field: The shear distribution in GREAT10 is a Gaussian
random field with random phases. In real data the field may have
non-Gaussian signatures, and nonrandom phase information.
\item
Known PSF: With real data we must estimate the PSF \textit{and} determine
the shear from each image. In GREAT10 we have separated these problems
into two challenges; in future GREAT challenges these aspects will be
combined.
\item
Weak Lensing: The shear field in GREAT10 only contains galaxies that
are weakly sheared (the local distortion entirely described by
$g_1$ and $g_2$). In reality, second-order effects are
present and galaxies can even be strongly distorted (into arcs) or
multiply imaged.
\item
Simple Galaxy Shapes: The galaxies used in this Challenge are simple
relative to real data (similar to GREAT08).
\item
Simple Noise Model: The noise in the GREAT10 images will contain a~Poisson
term from the image creation process, that mimic photon emission from
galaxies and stars,
and an additional Gaussian
component, to mimic noise in detectors. In practice, there are unusable
bad pixels which have to be flagged or removed.
\item
Background Estimation/Heteroskedasticity:
The noise in GREAT10 is constant across the image, where the data is
modeled as a signal plus additive noise. Real data will also have an uncertain
additive background component whose estimation further complicates
calculation of uncertainties.
\item
Image Construction: In GREAT10 objects are distributed across
the image with no overlaps. Furthermore, each object will be
classified to the participant as a star or galaxy. In reality, the
identification of objects is a~further problem, and objects commonly
overlap.
\item
Masking: The GREAT10 images do not have any data missing. In reality,
data can be missing or incomplete, for example, due to the layout of the
CCDs used to create the images. Furthermore, areas of images in real
data are
intentionally masked, for example, around very bright stars (that
saturate the
images) and satellite tracks (that leave bright linear trails across
images).
\item
Intrinsic Ellipticity: In GREAT10 the intrinsic
(nonsheared) ellipticities have only B-mode correlation (see Appendix
\ref{appC}). This is constructed to reduce the simulation size to a
manageable level. In real data the unsheared galaxies are expected to
have a random orientation (equal E- and B-mode). There are also
secondary effects that act to align galaxies that are in close
physical proximity which can contaminate the lensing
signal.\looseness=1
\end{itemize}
We envisage that the next GREAT challenge will build upon GREAT10 by
including one or more of these effects.

In addition, there are a multitude of further effects that will be
present in real data, for example, nonlinear CCD responsivity and
Charge Transfer Inefficiency (CTI), variation in
exposure times across a data set and variation of the PSF as a function
of wavelength, to name a few. A further challenge will be to handle
the enormous amount of data, of order petabytes, that will need to be
analyzed over reasonable timescales.

The ultimate challenge for methods developed and tested on the GREAT
challenges will be their application to data. In partnership with
state-of-the-art instrumentation, the GREAT challenges will help
scientists use gravitational lensing to answer some of our most
profound and fundamental questions
about the Universe.


\begin{appendix}\label{app}
\section{Removal of intrinsic shape noise in variable
shear simulations}\label{appA}

In shear simulations we must make some effort to reduce the effect of
the intrinsic ellipticity noise. This can be understood if we take an
example
requirement on the variance of shear systematics to be $\sigma^2_{
\mathrm{sys}}=10^{-7}$, which is a
requirement on the variance in the estimated shear values such that
our cosmological parameter estimation is unbiased.

Assuming that the intrinsic (unsheared) ellipticities are independent
and identically
distributed random variables, with a variance $\sigma^2_{\varepsilon}$, we
will need $N_{\mathrm{gal}}=
(\sigma^2_{\varepsilon}/\sigma^2_{\mathrm{sys}})$ to reach this
accuracy.\footnote{Since $|e|\leq1$ this argument is approximate,
since the distribution is truncated and hence not Gaussian.} If
$\sigma^2_{\varepsilon}\sim0.1$ (a typical empirically observed
quantity), this means $N\sim10^{6}$ galaxies are required
per shear value to reach the systematic floor. In GREAT08, with
$6\times50$ shear values over $9$ simulation conditions, this results
in a number which is large ($2.7\times10^8$ galaxies) and difficult
to analyze in a short timescale.

In constant shear simulations, as was done in GREAT08, galaxies can be
created in pairs such that they have the same unsheared ellipticity
except that one has been rotated by $90$ degrees. The $90$ degree
rotation converts an ellipticity to $e\rightarrow-e$.
If we have rotated (by 90 degrees) and unrotated galaxy pairs, then the
intrinsic shape noise cancels to first order and the variance on the
shear is reduced to $g^2 (\sigma^4_{\varepsilon}/2N)$ [\citet
{2007MNRAS37613M}]. This can be understood if we
write the shear estimator for a single object $i$ as
%
%
\begin{equation}
\label{rot1}
\tilde g^i_{\alpha}=e^i_{\alpha}+P^{i}g^i+{\mathcal O}(g^2),
\end{equation}
where $P$ is some response factor (a matrix that partially
encodes the effect of the PSF) and higher-order terms also encapsulate
any noise. In a $90$ degree rotated image the shear estimator can be
written as
%
%
\begin{equation}
\label{rot2}
\tilde g^i_{\alpha}=-e^i_{\alpha}+P^{i}g^i+{\mathcal
O}(g^2).
\end{equation}
It can be easily seen that by averaging the individual shear
estimates the intrinsic ellipticity contribution cancels.

In variable shear simulations, like GREAT10, the $90$ degree rotation
me\-thod cannot be used. In GREAT10 we require the E-mode correlation
function or power spectrum to be reconstructed. If we have two images
with $90$ degree rotated galaxies, then by taking the correlation
of the above shear estimates equations (\ref{rot1}) and (\ref{rot2})
we have
%
%
\begin{eqnarray}
\label{rot3}
\langle\tilde g_{\alpha}\tilde g_{\beta}\rangle_{\mathrm{unrot}}&=&
\langle
e_{\alpha}e_{\beta}\rangle+P^2\langle g g\rangle+{\mathcal
O}( g^4),\\
\langle\tilde g_{\alpha}\tilde g_{\beta}\rangle_{\mathrm{rot}}&=&
\langle e_{\alpha}e_{\beta}\rangle+P^2\langle g g\rangle+{\mathcal O}( g^4).
\end{eqnarray}
It can be seen from these that we cannot combine correlation function
estimates from rotated and unrotated images to remove the intrinsic
ellipticity correlation function, and preserving the shear correlation
function. This is fundamentally because the intrinsic ellipticity
correlation function is a random field, with equal E- and B-mode
contributions.

For variable shear simulations we can take advantage of the fact that
the shear correlation function contains E-mode correlations only.
In GREAT10 the intrinsic ellipticity correlation function will
contain B-mode correlations only. This means that the correlation
function from any image can be written as in (\ref{rot3}),
where now $\langle e_{\alpha}e_{\beta}\rangle$ only contains B-mode
correlation and $\langle g g\rangle$ only contains E-mode
correlation. By taking the E-mode component of the shear estimate
$\langle\tilde g\tilde g\rangle$, we
will then eliminate any contribution from the intrinsic ellipticity
distribution. Thus, the simulation size, like the $90$ degree-rotated
case for constant shear images, will not be determined by the intrinsic
ellipticity variance.

This step is unrealistic in that real galaxies are not expected to
have any preferred unsheared correlation (to first order),
but is necessary for public
simulations to make the size of the challenge approachable.

\vspace*{6pt}
\section{Calculating the shear correlation function and
shear power spectrum for GREAT10}\label{appB}

In this Appendix we present methods that can be used to
estimate the shear correlation function and power spectrum from the
individual shear estimators
from each galaxy.

At the beginning of the challenge we will
provide links to open-source code to calculate these
various statistics, given a catalogue of shear values per galaxy. The
Galaxy Challenge submission procedure will allow for either a shear
catalogue (per galaxy), a correlation function or a power spectrum to
be submitted.

\subsection*{Calculating the correlation function}

Here we summarize Crittenden et al. (\citeyear{2002ApJ56820C}) (and reference to this
article) where shear-correlation statistics are presented in detail.

If we
define a Cartesian coordinate system ($\theta_x$, $\theta_y$) on a
patch of sky (assumed small) and $\phi$ is the angle between the
line joining two galaxies and the $x$-axis, then the
tangential component can be written as $g_{+}=-[g_1\cos(2\phi)+g_2\sin
(2\phi)]$
and the cross term as $g_{\times}=-[g_1\sin(2\phi)-g_2\cos(2\phi)]$.
We can take the correlation function of these two quantities
%
%
\begin{eqnarray}
\label{eq2}
\xi(\Delta\thetab)_{+}&=&\langle g_{+}(\theta_x,\theta_y)g_{+}(\theta
_x,\theta_y;\Delta\thetab)\rangle,\nonumber\\[-8pt]\\[-8pt]
\xi(\Delta\thetab)_{\times}&=&\langle g_{\times}(\theta_x,\theta
_y)g_{\times}(\theta_x,\theta_y;\Delta\thetab)\rangle,\nonumber
\end{eqnarray}
where $\Delta\thetab$ is the average radial (angular) separation of galaxy
pairs, and the angle brackets represent an average over all
galaxy pairs within a considered~$\theta$.
Here we refer to sums over finite-width bins in $\theta$.

The E- and B-mode correlation functions are related to these
observable correlations, combinations of
$\xi_+(\Delta\thetab)$ and $\xi_-(\Delta\thetab)$, in the following
way:
%
%
\begin{eqnarray}
\xi_E(\Delta\thetab)&=&\tfrac{1}{2}[\xi_+(\Delta\thetab)+\xi_{\times
}(\Delta\thetab)]
+\tfrac{1}{2}\nabla^4\chi^{-1}[\xi_+(\Delta\thetab)+\xi_{\times}(\Delta
\thetab)],\nonumber\\[-8pt]\\[-8pt]
\xi_B(\Delta\thetab)&=&\tfrac{1}{2}[\xi_+(\Delta\thetab)+\xi_{\times
}(\Delta\thetab)]
-\tfrac{1}{2}\nabla^4\chi^{-1}[\xi_+(\Delta\thetab)-\xi_{\times}(\Delta
\thetab)].\nonumber
\end{eqnarray}
The operator $\nabla^4\chi^{-1}$ is most easily evaluated in
Fourier space
%
%
\begin{equation}
\nabla^4\chi^{-1}g(r)=\int r'{\mathrm d}r' g(r'){\mathcal G}(r,r'),
\end{equation}
where
%
%
\begin{equation}
{\mathcal G}(r,r')=\int\frac{k\,{\mathrm d}k}{2\pi}J_0(kr)J_4(kr').
\end{equation}
See Crittenden et al. (\citeyear{2002ApJ56820C}) for more information. In
Appendix \ref{appE} we
define the correlation function in a complementary way in
\eqref{s02corr}, where the binning in $\theta$ is more explicit.

Note that we expect the cosmological signal to contain nearly zero
B-mode correlations (see Figure \ref{varshear}), but that in GREAT10
the B-modes reveal the intrinsic ellipticity distribution by design
(see Appendix \ref{appA}). In GREAT10 the
quality factor will be evaluated using the E-mode correlation only.

\subsection*{Calculating the power spectrum}

Recall from (\ref{eq1}) that we have two shear components
and that these vary as a function of position across the fields
$g_1(\theta_x,\theta_y)$ and $g_2(\theta_x,\theta_y)$.
We can write the ``shear'' as a complex number such that
%
%
\begin{equation}
g=g_1+{\mathrm i}g_2,
\end{equation}
and we can Fourier transform the shear in the following way
%
%
\begin{equation}
g_R(\ell_x,\ell_y)+{\mathrm i}g_I(\ell_x,\ell_y)=\sum_{
\mathrm{galaxies}}[g_1(\theta_x,\theta_y)+{\mathrm i}g_2(\theta_x,\theta_y)]
{\mathrm e}^{-{\mathrm i}\bell.\thetab},
\end{equation}
where two new Fourier estimators have been created, a real and an
imaginary part which are a function of $\ell=(\ell_x$, $\ell_y)$. These
are simply related to the original $g_1$ and $g_2$ by
%
%
\begin{eqnarray}
g_R(\ell_x,\ell_y)&=&\sum_{
\mathrm{galaxies}}g_1(\theta_x,\theta_y)\cos(\ell_x\theta_x+\ell_y\theta
_y)\nonumber\\
&&{}+g_2(\theta_x,\theta_y)\sin(\ell_x\theta_x+\ell_y\theta_y),\nonumber
\\[-8pt]\\[-8pt]
g_I(\ell_x,\ell_y)&=&\sum_{
\mathrm{galaxies}}g_2(\theta_x,\theta_y)\cos(\ell_x\theta_x+\ell_y\theta
_y)\nonumber\\
&&{}-g_1(\theta_x,\theta_y)\sin(\ell_x\theta_x+\ell_y\theta_y).\nonumber
\end{eqnarray}
We now have two estimators that are a function of $\ell_x$ and
$\ell_y$, with each point in the ($\ell_x$, $\ell_y$) plane being
a sum over all galaxies.

In cosmology we wish to decompose the shear field into
an E- and a~B-mode; cosmological structures should only create an
E-mode signal.
To make this decomposition, we can describe the
($\ell_x$, $\ell_y$) by an angle $\phi_{\ell}=\tan^{-1}(\ell_y/\ell_x)$
and a
scalar $\ell=\sqrt{\ell_x^2+\ell_y^2}$. The E- and B-mode fields can
now be written as a simple rotation of the Fourier plane
%
%
\begin{eqnarray}
E(\ell_x,\ell_y)+{\mathrm i}B(\ell_x,\ell_y)&=&
[\cos(2\phi_{\ell})-{\mathrm i}\sin(2\phi_{\ell})]\nonumber\\
&&{}\times[g_R(\ell_x,\ell_y)+{\mathrm i}g_I(\ell_x,\ell
_y)],\nonumber\\[-8pt]\\[-8pt]
E(\ell_x,\ell_y)&=&\cos(2\phi_{\ell})g_R(\ell_x,\ell_y)+\sin(2\phi_{\ell
})g_I(\ell_x,\ell_y),\nonumber\\
B(\ell_x,\ell_y)&=&\cos(2\phi_{\ell})g_I(\ell_x,\ell_y)-\sin(2\phi_{\ell
})g_R(\ell_x,\ell_y).\nonumber
\end{eqnarray}
Finally, the shear power spectrum can be defined as the modulus of
the E-mode field
%
%
\begin{equation}
C(\ell)=\frac{|E(\ell_x,\ell_y)|^2}{2\pi},
\end{equation}
where in practice the $C(\ell)$ is the average over some bin in $|\ell|$.

For data that is distributed as a grid in $\theta_x$ and $\theta_y$,
the above calculations can be simplified even further written as a
series of FFTs:
\begin{itemize}
\item
Make a 2D FFT of the shear field
$ g(\theta_x,\theta_y)\rightarrow\mathcal{G}(\ell_x,\ell_y)$.
\item
Construct a 2D $\ell$-matrix $\bell=(2\pi/\theta_x)+{\mathrm i}(2\pi
/\theta_y)$.
\item
Rotate the FFT of the shear field
$\mathcal{G}^R(\ell_x,\ell_y)\rightarrow(\bell^*\bell^*/|\bell
^2|)\mathcal{G}(\ell_x,\ell_y)$.
\item
Inverse FFT the rotated shear field back to real space
$\mathcal{G}^R(\ell_x,\ell_y)\rightarrow E(\theta_x,\allowbreak\theta_y)+{\mathrm
i}B(\theta_x,\theta_y)$.
\item
Select the real part, E-mode and FFT $E(\theta_x,\theta_y)\rightarrow
\mathcal{E}(\ell_x,\ell_y)$ (which is now complex).
\item
Calculate the modulus
$|\mathcal{E}(\ell_x,\ell_y)|^2=
\mathrm{Re}[\mathcal{E}(\ell_x,\ell_y)^2]+
\operatorname{Im}[\mathcal{E}(\ell_x,\ell_y)^2]$
and azimuthally bin in $|\bell|$ to find the power spectrum.
\end{itemize}

At the launch of GREAT10 we will specify exactly which
binning scheme in $\ell_x$, $\ell_y$ and $\ell$ we will use to
calculate the result, and we will also specify the angular binning
$\theta$ we require for correlation function submissions.

Note that the method described in this section
will not work in the case that images are
masked. In Section \ref{GREAT10 Simplifications and Future Challenges}
we describe how masking is present in real data
and may be included in future GREAT challenges.

\section{Overview of existing shape measurement~methods}\label{appC}

We refer the reader to the GREAT08 articles, Bridle et al.
(\citeyear{2009AnApS36B}), for a comprehensive
review of
shape measurement methods up until late 2009.
Since the publication of the GREAT08 results paper and up until\vadjust{\eject}
the creation of this Handbook, mid-2010,
there have been several ongoing activities in the field.

The GREAT08 simulation details have been made public, both shear
values (answers) for the simulations have been made available and the
properties of each individual object in the simulations (these can be
found by following the relevant links from
\url{http://www.greatchallenges.info}).

In the interim period a new blind
realization of GREAT08 has been released, GREAT08reloaded. This
simulation is \textit{exactly} the same as GREAT08 except that the true
shears (answers) have new values. These simulations can be used as a
semi-blind challenge such that developing algorithms can be tested
on new simulated data. A new online quality factor calculator is
available,
but we do not show an online ranking at
\texttt{\href{http://www.greatchallenges.info}{http://www.greatchallenges.}
\href{http://www.greatchallenges.info}{info}}. Authors of newly developed
algorithms are
encouraged to publish new quality factor results from these simulations
(please inform T. Kitching and S. Bridle if you intend to publish
results based on GREAT08reloaded).

Some new methods have been published in this interim period.
\citet{2010arXiv10012333B} has presented an investigation into
several effects, related
to model fitting algorithms, that
impact shear measurement biases. \citet{2010arXiv10023615Z} has
investigated mapping from
ellipticity to shear and has presented a number of shear
estimators with varying success.
\citet{2010arXiv10020838G} have used a neural
net approach, similar to those used for estimating galaxy distances
(redshifts), to estimate shear. \citet{2010arXiv10081076M} have
introduced a model-independent deconvolution method.
Two of these new approaches, \citet{2010arXiv10012333B} and
\citet{2010arXiv10020838G},
have claimed significantly improved
GREAT08 quality factors of $Q_{\mathrm{GREAT08}}\gs1\mbox{,}000$ on the
low-noise
subsets, and \citet{2010arXiv10081076M} have claimed $Q_{\mathrm
{GREAT08}}\gs
500$ on some real-noise subsets.

\section{Overview of existing PSF modeling methods}\label{appD}

In this Appendix we will describe the techniques currently used by the
gravitational lensing community to characterize the spatial variation
of the PSF. We classify these approaches under three broad headings, as
discussed in Section \ref{Variable PSF}.

\subsection*{Direct modeling}

The modeling of the spatial variation of the PSF across an image
can take the form of fitting simple, continuous surfaces to
quantities that parametrize the PSF at a given location. In this case
a surface is fit to the image and the best fit is found by determining an
extreme value of some goodness of fit, usually a minimum chi-square,
calculated discretely at the star positions. The quantity being fit
can either be the pixel intensities themselves or some derived
quantity such as the ellipticity and Full-Width-Half-Maximum (FWHM)
of the stars.

For example, in the KSB
shear measurement method (see the GREAT08 Handbook Appendix B) these
are usually the two components of the stellar anisotropy
correction, estimated using the measured ellipticities of stars.
Another method is to model PSFs using shapelet basis
functions [e.g., \citet{2002AJ123583B};
\citet{2003ARAA41645R};
\citet{2005MNRAS363197M}; \citet{2006astroph10606K};
\citet{2008MNRAS385695B}],
where at each star a shapelet model is fit and
it is the spatial variation in each shapelet coefficient that
describes
the PSF surface. In \textit{lens}fit
[\citet{2007MNRAS382315M}; \citet{2008MNRAS390149K}]
implementations to date,
the PSF is modeled on a
pixel-by-pixel basis where for each pixel in a postage stamp a
2D polynomial is fit across the image. PSFEx [e.g., \citet
{2003AJ1261402K}] combines several model fitting aspects, allowing for various
different orthonormal expansions in 2D to be used.
Regardless of the PSF model parameter being modeled, there is
considerable freedom in the choice of the functional form of the
fitting surface [see, however, Rhodes et al.
(\citeyear{2007ApJS172203R}) and \citet{2008arXiv08100027J}
for attempts to use PSF models motivated by realistic
optical patterns]; however, simple bivariate polynomials are typically used.

There are well-known problems with polynomial fitting surfaces,
including reduced
stability at field edges and corners, as the fits become poorly
constrained. These have been noted but not
necessarily tackled beyond suggestions of other, perhaps better
behaved, functional schemes [e.g., \citet{2005AA42975V}].
Alternatively, it has been suggested [e.g., \citet{2006astroph10606K}]
that the images can be Gaussianized to create a better behaved
local functional variation of the PSF, at expense of correlated noise
properties,
however, global PSF interpolation is still required for such processes.

\subsection*{Indirect modeling: Principal component analysis (PCA)}

Indirect modeling applies to a class of methods that do not model the PSF
explicitly, for example, using some kind of polynomial, but attempt to
characterize the variation across the image by finding patterns that are
present between many realizations of the same underlying PSF.
Principal Component Analyses (PCA; also known as eigenfunction
analyses) are an example of this kind of PSF modeling
that has been implemented by the community.

The motivation for such techniques is that gravitational lensing
images (especially at particular positions, for example, away from the
plane of
the Milky Way) often contain only $\ls$1
star/arcmin$^2$ usable for PSF measurements.
If the PSF shows spatial variations on this or smaller scales,
they cannot be modeled reliably using a standard direct modeling
interpolation, for example, with polynomial functions.
However, these small-scale variations may show some degree of stability
between different images (exposures) obtained using the same
instrument, depending on their physical origin.
If this is the case, one can attempt to combine the information
from different exposures---with stars at different pixel
positions---to obtain a PSF model with improved spatial interpolation.

A PCA interpolation was suggested for use in gravitational lensing
by \citet{2004astroph12234J}.
They used PCA to identify the main directions
of PSF variation between their exposures. This is typically done by first
computing a standard polynomial interpolation in each exposure,
where the polynomial order should be chosen to be
sufficiently low to avoid overfitting.
The PCA is then run on these \textit{polynomial coefficients}.
In this way the recurring PSF patterns in the different exposures can be
described as a~superposition of principal component patterns, where
each exposure is characterized by its principal component coefficients.
These coefficients yield a~sorting scheme which enables the
desired combination of stars from different exposures.

As the final step the stars from all exposures are fit together
with a high-resolution model.
In the description of \citet{2004astroph12234J} this model
contains one
higher-order spatial polynomial
for each principal component.
For the stars of a certain exposure these different polynomial terms
are weighted according to the
principal component coefficients of this exposure.
Besides the polynomial order, one has to choose the number of included
principal components. The first principal component
is the most important one, carrying the largest variation in PSF space.
The higher principal components carry less and less variation, and may at
some point be truncated to avoid the fitting of noise, for example,
once 99\% of the
variation are taken into account.

Some of the principal components (eigenfunctions) can have a physical
interpretation, for example, in relation to the telescope optics (see
Section~\ref{Variable PSF}).
The principal components of the PSF
are expected to have some relation to the main physical effects
influencing the PSF,
such as changes in telescope focus, seeing, wind-shake or pointing elevation.
However, in detail, the cross-identification may be difficult,
in particular, for the higher, less important principal components.

The PCA approach does not capture PSF variations
which are completely random and appear in a single exposure only.
On large scales these can be accounted for by combining the PCA model
with a low-order residual polynomial fit, computed separately for each
exposure [see Appendix B.5 of \citet{2010AA516A63S}].
However, on small scales such random variations cannot be corrected.

Also, given that PCA is a linear coordinate transformation, it does not
efficiently
describe possible nonlinear distributions of exposures in the PSF coefficient
space. Such nonlinear distributions may, in particular, occur if
PSF quantities with different responses to physical parameters are fitted
together. For example, telescope defocus changes PSF ellipticity
linearly, whereas PSF size is affected quadratically.
If one aims at a compact PSF description with few principal components,
this can partially be compensated using additional terms in the final
PCA fitting step,
which depend nonlinearly on the principal component coefficients [see
\citet{2010AA516A63S}].

\subsection*{Image deconvolution}

Astronomical images can be blurred for various reasons:
for example, atmospheric turbulence for ground-based telescopes, or thermal
deformations of the telescope through varying exposure to sunlight for
space-based telescopes. Methods for deconvolution and deblurring can
often alleviate those effects
[e.g., \citet{kundur} and references therein].
There are three aspects of the deconvolution process that we comment
on here:
\begin{itemize}
\item
Nonblind vs. blind. In astronomy the point spread function
(PSF), which describes the blur, can often be estimated from nearby
stars, allowing \textit{nonblind} deconvolution to deblur recorded
images (called nonblind because the PSF is known). However, the PSF
might not be available or cannot be reliably guessed. In that case
\textit{blind} methods try to recover simultaneously the PSF and the
underlying image. This is usually achieved by optimization with
alternating projections in combination with prior assumptions
about the PSF and the image.
\item
Single-frame vs. multi-frame. Especially important for
nonastronomical photography is single-frame deconvolution, in which
we only observe a~single blurry image and would like to recover a
deblurred image [e.g., \citet{Fergus2006} and \citet
{Cho2009}]. In
astronomy we
are often able to record multiple images of the same object.
Combining a large number of frames can then recover a single deblurred
frame [e.g., by speckle-interferometric methods such as
\citet{labeyrie}; \citet{knox}, or by multi-frame
deconvolution methods,
e.g., \citet{schulz}; \citet{harmeling09}].
\item
Space-invariant vs. space-variant. A common assumption is that
the PSF is space-invariant, that is, in different parts of the image, the
blur of a~single star looks the same. For atmospheric turbulence this
assumption holds only inside the \textit{isoplanatic patch}, which is a
small angular region the size of which depends on the amount of air
turbulence. For images of larger angular regions (beyond the
isoplanatic patch), the PSF does change across the image and this must
be taken into account. Common approaches chop the image into
overlapping patches with constant PSF and interpolate between them
[e.g., \citet{Nagy1998}; \citet{Bardsley2006}; \citet{hirsch10}].
\end{itemize}

Finally, we note that in astronomical images the presence of
significant noise (typical objects used for cosmic shear have
signal-to-noise ratios of $\approx$10--$20$) is a challenge to
deconvolution algorithms.

\subsection*{Other methods}

While the direct and indirect modeling methods described so far in
this Appendix have been fully tested and applied to real data, there
are other techniques that have been proposed as potentially useful for
modeling the spatial variation in the PSF. We now discuss two
of these suggestions in brief.

Kernel Principal Component Analysis [KPCA: e.g., \citet{SSM98};
\citet{ecs9580}], a highly successful tool
in pattern recognition, has been suggested as a further aid to
stable PSF modeling.
Variation in even basic aspects of a PSF (overall size, ellipticity,
orientation) may often be difficult to succinctly describe in terms of
linear combinations of image pixel values. This means that PCA models
may sometimes require unnecessary degrees of freedom, which impacts
upon the stability of accurate shear measurement. KPCA allows the
Principal Components to be separated in a higher-dimensional feature
space, mapped implicitly via a dot-product kernel, where these
nonlinear dependencies can be unraveled.

Another statistical tool that might prove useful in spatial modeling
of the PSF is known as kriging, commonly employed in the field of
Geostatistics [see, e.g., \citet{Cressie}].
This technique uses a weighted average of neighboring samples (in
this case stars) to estimate the value of an unknown function at a
given location. Weights are optimized using the location of the
samples and all the relevant interrelationships between known and
unknown values. The technique also quantifies confidence in model
values. However, the accuracy of the method relies critically upon
assumptions about the nature of the underlying variation;
commonly-used types of kriging (such as Ordinary, Universal and
IRFk-kriging) reflect the impact of differing choices for these
assumptions. A further complication is the limited work done toward
the application of kriging in the presence of noisy sampled data.

\section{Star challenge quality factors}\label{appE}

The estimation of the PSF in images has a broad range of
applications in astronomy beyond gravitational lensing, hence, we will
define a
quality factor that is flexible enough to allow nonlensing applications
to gain
useful information.

Each participant will submit a high resolution estimate of the PSF at
each nonstar position that is requested. The quality factor that we
will use to estimate the success of a method at estimating the PSF
will calculate the residual (true PSF minus submitted estimate) for
each nonstar
position for each pixel. For astronomical analyses we are
concerned with two basic parameters that have the highest
impact on shape of the PSF:
these are the ellipticity $e$ of the PSF and
the size $R$ of the PSF [\citet{2008AA48467P}]. These are
defined as follows [see \citet{2001PhR340291B}, Section 4.2].

First we define the second-order brightness moments of the image as
%
%
\begin{equation}
q_{ij}=\frac{\sum_p w_p I_p (\theta_i - \bar\theta_i)(\theta_j - \bar
\theta_j)}{\sum_p
w_p I_p},\qquad i,j\in\{1,2\},
\end{equation}
where the sums are over pixels, $I_p$ is the flux in the $p$th
pixel and $\theta$ is a pixel position ($\theta_1=x_p$ and $\theta
_2=y_p$). We
include a weight function $w_p$, that will be defined when the
simulations are launched, to ensure that the sums over pixels converge
for the exact PSF model used. We now write the weighted ellipticity
for a PSF in complex notation as
%
%
\begin{equation}
e_{\mathrm{PSF}} = \frac{q_{11}^2 - q_{22}^2 + 2{\mathrm
i}q_{12}}{q_{11}^2 +
q_{22}^2 + 2(q_{11}q_{22}-q^2_{12})^{1/2}},
\end{equation}
where we have used a definition of ellipticity $|e|=(1-r)(1+r)^{-1}$
which is consistent with Section \ref{The Simulations}; note there is
an equivalent expression for $|e|=(1-r^2)(1+r^2)^{-1}$
[see \citet{2001PhR340291B}, Section 4.2].
For the weighted size we have a similar expression
%
%
\begin{equation}
R_{\mathrm{PSF}}^{2} = q_{11} + q_{22}.
\end{equation}
We can calculate the variance between the ellipticity of the model
and true PSF $\sigma^2(e_{\mathrm{PSF}})\equiv\langle(e_{\mathrm{PSF}}
- e_{
\mathrm{PSF}}^{\mathrm{t}})^2\rangle$ and similarly for the size
$\sigma^2(R_{\mathrm{PSF}})\equiv\langle(R_{\mathrm{PSF}} - R_{
\mathrm{PSF}}^{\mathrm{t}})^2\rangle$.

For cosmic
shear studies we have the requirement that we need the residual error
in the ellipticity and size to be $\ls$10$^{-3}$ for the impact on
cosmological parameter estimation to be low;
for more detail see \citet{2008AA48467P}, \citet
{2009AA500647P}. Hence, we
define the quality factor for the Star Challenge as
%
%
\begin{equation}
P\equiv\frac{1}{({1/2})\langle\sigma^2(R_{\mathrm{PSF}})/R^2+\sigma
^2(e_{\mathrm{PSF}})\rangle},
\end{equation}
where the angle brackets represent an average over all images.
There are further steps
that could be taken that can map PSF residuals onto cosmic shear
requirements [and indeed the Galaxy Challenge $Q$ factor;
\citet{2008AA48467P}]; we will present these in the final GREAT10
analysis.

In the following sections
we outline more details on some additional PSF quality factors that we will
employ in the final analysis. As well as the summed residual between
the model and the data, we will investigate at least two other
approaches that are more closely matched to the gravitational lensing
requirements (but are less applicable to other fields of
astronomy).

These techniques will not be used to determine the outcome of the
competition or used to create live quality factors during the
challenge.

\subsection*{Azithumal statistics}
We can quantify the quality of the submitted PSF model in terms of how
well a galaxy shear is recovered if this model were used. To do this,\vadjust{\eject}
we will convolve the true PSF with a simple galaxy model, and then use the
submitted PSF to recover the galaxy parameters from the
true PSF-convolved image. This can be done for several angles in an
azimuthal bin (ring) (and
be repeated for zero shear and nonzero shear) to obtain
multiplicative and additive errors on the measured shear [$m$ and~$c$;
see the Shear TEsting Programme; \citet{2006MNRAS3681323H};
\citet{2007MNRAS37613M}].
These $m$, $c$ values can be converted into
a GREAT08 $Q$ value (assuming
a constant local shear).
By comparing this ring test with the summed residuals, we will
be able to correlate the magnitude of the residuals
to the bias on the shear (i.e., $Q$).


\subsection*{Autocorrelations in residuals}

The autocorrelation of any continuous function $h$ across its
$n$-dimensional domain can be defined as
%
%
\begin{equation}\label{eq:fstarg}
[h \star h](\rb) \equiv\int h^*(\xb) h(\xb+ \rb) \,\dif^n x.
\end{equation}
We may also employ the function $h(\xb)$ to represent a discrete,
noisy sampling of an unknown ``true'' field $h_{\mathrm{t}}(\xb)$ such that
%
%
\begin{equation}
h(\xb) = h_{\mathrm{t}}(\xb) + N(\xb),
\end{equation}
where $N$ is a stochastic noise term. This description of discrete
observations by a continuous ``quasi-field'' is a convenient notational
shorthand in what follows; one can imagine the observations as a
smooth field convolved with delta functions at the data locations. The
data represented by $h(\xb)$ can be pixel values, complex
ellipticities or any general data vector.
Similarly, a~best-fitting model used to characterize these
observations can be written~as
%
%
\begin{equation}
h_{\mm}(\xb) = h_{\mathrm{t}}(\xb) + m(\xb, h_{\mathrm{t}}, N ; f_{\mm}),
\end{equation}
where the $m = h - h_{\mathrm{t}}$ is referred to as the
\textit{inaccuracy} in the model [\citet{2010MNRAS404350R}] and
$f_{\mm}$ labels the
specific modeling scheme chosen to represent $h(\xb)$.

The unknown function $m$ will depend nontrivially upon the data,
noise and modeling scheme used, but not all of its properties are
entirely hidden.
If we make two simple assumptions about the noise in the data being
considered, that
%
%
\begin{equation}\label{eq:noise1}
[N \star N](\rb) = 0
\end{equation}
and
%
%
\begin{equation}\label{eq:noise2}
[h_{\mathrm{t}} \star N + N \star h_{\mathrm{t}} ](\rb) = 0,
\end{equation}
it follows that
%
%
\begin{equation}\label{eq:gcorr}
[ (h-h_{\mm}) \star(h-h_{\mm}) ] (\rb) = [m
\star m ](\rb) - [m \star N + N \star m ](\rb)
.
\end{equation}
As demonstrated in \citet{2010MNRAS404350R}, this residual
autocorrelation function
should tend to zero for all $\rb$ if the fitting model employed is
both stable and accura\-te: the residuals are then
indistinguishable
from pure white noise. Departures from this ideal behavior have\vadjust{\eject}
predictable consequences: overfitting models will show consistently
negative values of $ (h-h_{\mm}) \star(h-h_{\mm})$ (since $[m
\star N + N \star m ] > m \star m > 0$ in these cases),
whereas underfitting models (for which $m \star m$ dominates) may be
positive or negative in general.

If we consider the data in question to be an array of image pixels
$I_i(\thetab_i)$, with a corresponding best-fitting model $I_{\mm}$,
we can construct a practical estimator $\hat{\xi}_I$ for the
autocorrelation function in \eqref{eq:gcorr}. We do this
only in the isotropic case $\hat{\xi}_I(\theta) =
\hat{\xi}_I(\thetab)$ for simplicity, but note that the
autocorrelation may not be isotropic in general.

Following \citet{2002AA3961S}, we estimate the
correlation function in bins of finite width $\Delta\theta$ and
define the function $\Delta_{\theta} (\phi) = 1$ for $\theta- \Delta
\theta/ 2 < \phi< \theta+ \Delta{\theta} / 2$ and zero otherwise:
thus, $\Delta_{\theta}(\phi)$ defines the bin at separation~$\theta$.
A~simple estimator $\hat{\xi}_I$ for $(I-I_{\mm}) \star(I-I_{\mm})$
may then be written as
%
%
\begin{equation}
\label{s02corr}
\hat{\xi}_I(\theta) = \frac{1}{N_{\mathrm{p}} (\theta) }\sum_{ij} w_i
w_j [I_i(\thetab_i) -I_{\mm}(\thetab_i) ] [
I_j(\thetab_j) - I_{\mm}(\thetab_j) ] \Delta_{\theta} (
|\thetab_i - \thetab_j| ) ,\hspace*{-28pt}
\end{equation}
where the sum is over all pixel pairs, the weight $w_i$ assigned to
the $i$th pixel may be used to account for variations in the
signal-to-noise across the image plane, and $N_{\mathrm{p}}(\theta) =
\sum_{ij} w_i w_j \Delta_{\theta} ( |\thetab_i - \thetab_j
|)$ is the effective number of pairs in the bin considered.

\section{Rules of the challenge}\label{appF}

The challenge will be run as a competition, and the winner will be
awarded with a prize.

We will
award a prize to the participants with the highest quality factor in
the Galaxy Challenge at the end of the submission period. Prizes will be
awarded at a final GREAT10 meeting, and
winners will be required to make a descriptive presentation of their
method at this workshop.

To define the scope of the competition we
outline some participant rules here:

\begin{longlist}[(1)]
\item[(1)]
Participants may use a pseudonym or team name on the results leader
board, however, real names (as used in publications) must be provided
where requested during the result submission process. We will also require
an email address to be provided, so that we can communicate GREAT10 information
directly to participants. Participant details will be confidential, and
no information
will be made available to any third party.
\item[(2)]
Participants who have investigated several algorithms may enter once per
method. Changes in algorithm parameters do not constitute a different
method.
\item[(3)]
Resubmissions for a given method may be sent a maximum of once per
week per challenge during the 9 month competition. There will be
allowed 1
submission per week for: Star Challenge and Galaxy Challenge.
If submission rates fluctuate, the
submission time interval may be altered to accommodate the needs of
the participants.\vadjust{\eject}
\item[(4)]
Participants must provide a report detailing the method used, at the
Challenge deadline. We
will also provide a webpage where we will encourage participants to
keep a log of their activities. If participants would like to provide
code, this can also be uploaded to the webpage.
\item[(5)]
Any publication that contains results relating to the GREAT10
simulations, that authors wish to submit during the 9 months of the
challenge, must be approved
by the GREAT10 PI (T. D. Kitching; tdk@roe.ac.uk)
and GREAT10 Advisory Team before submission
to any journal or online archive.
\item[(6)]
We expect all participants to allow their results to be included in the final
Challenge Report. We will, however, be flexible in cases where methods
performed badly if participants are against publicizing them.
We will release the true shears and PSFs (and variation in power
spectrum)
after the deadline.
\item[(7)]
Participants are encouraged to
freely write research articles using the Challenge simulations, after the
submission of the GREAT10 results article. We especially encourage
participants to submit articles on thier methods to the host journal
for GREAT10 \textit{Annals of Applied Statistics}.
\item[(8)]
The simulations may be updated during the challenge and/or modified,
if any improvements are required.
If any modification occurs,
participants will be notified by the email addresses provided at
submission, and any changes will be posted
on the GREAT10 website.
\end{longlist}

\subsection*{GREAT10 team rules}

The GREAT10 Team is defined as the PI (T. D. Kitching),
a GREAT10 Coordination Team whose role is
to make decisions
related to the input properties of the simulations, and a GREAT10
Advisory Team whose role is to advise on all other matters that do not
directly influence the simulations (e.g., workshop activities).

Some additional competition rules apply to members of the GREAT10
Coordination Team and PI:
\begin{longlist}[(1)]
\item[(1)]
For the purpose of these rules, the GREAT10 Coordination Team is defined
as being anyone who has\vadjust{\goodbreak} participated in a GREAT10 Coordination Team
meeting (in person, video or phone conference) or who has access to the
GREAT10 Coordination website.
\item[(2)]
Only information available to non-GREAT10 participants may be used
in carrying out the analysis, for example, no inside information about the
setup of the simulations may be used.
Note that the true shear values will only be available to
an even smaller subset of the GREAT10 Coordination Team and PI.
\item[(3)]
Any submission from the GREAT10 Coordination Team, or PI,
will be highlighted in the GREAT10 results article, in a similar way to
GREAT08 Team members in \citet{2009AnApS36B}.
\end{longlist}
The above rules do not apply to the members of the GREAT10 Team who are
in the GREAT10 Advisory Team only.
\end{appendix}

\section*{Acknowledgments}

We especially thank Lance Miller,
Fergus Simpson, Antony Lewis and Alexandre Refregier for useful discussions.
We acknowledge the EU FP7 PASCAL Network for funding this project, and
for electing GREAT10 to be a 2010 PASCAL Challenge. In particular, we
thank Michele Sebag and Chris Williams for their help.




%
\printaddresses


\begin{thebibliography}{48}

\bibitem[\protect\citeauthoryear{Amara and
  R{\'e}fr{\'e}gier}{2008}]{2008MNRAS391228A}
\begin{barticle}[author]
\bauthor{\bsnm{Amara},~\bfnm{A.}\binits{A.}} \AND
  \bauthor{\bsnm{R{\'e}fr{\'e}gier},~\bfnm{A.}\binits{A.}}
(\byear{2008}).
\btitle{Systematic bias in cosmic shear: Extending the Fisher matrix}.
\bjournal{Monthly Notices of the RAS}
\bvolume{391}
\bpages{228--236}.
\end{barticle}
\endbibitem

\bibitem[\protect\citeauthoryear{Bardsley et al.}{2006}]{Bardsley2006}
\begin{barticle}[author]
\bauthor{\bsnm{Bardsley},~\bfnm{J.}\binits{J.}},
  \bauthor{\bsnm{Jeffries},~\bfnm{S.}\binits{S.}},
  \bauthor{\bsnm{Nagy},~\bfnm{J.}\binits{J.}} \AND
  \bauthor{\bsnm{Plemmons},~\bfnm{B.}\binits{B.}}
(\byear{2006}).
\btitle{A computational method for the restoration of images with an unknown,
  spatially-varying blur}.
\bjournal{Optics Express}
\bvolume{14}
\bpages{1767--1782}.
\end{barticle}
\endbibitem

\bibitem[\protect\citeauthoryear{Bartelmann and
  Schneider}{2001}]{2001PhR340291B}
\begin{barticle}[author]
\bauthor{\bsnm{Bartelmann},~\bfnm{M.}\binits{M.}} \AND
  \bauthor{\bsnm{Schneider},~\bfnm{P.}\binits{P.}}
(\byear{2001}).
\btitle{Weak gravitational lensing}.
\bjournal{Physrep.}
\bvolume{340}
\bpages{291--472}.
\end{barticle}
\endbibitem

\bibitem[\protect\citeauthoryear{Berg{\'e} et al.}{2008}]{2008MNRAS385695B}
\begin{barticle}[author]
\bauthor{\bsnm{Berg{\'e}},~\bfnm{J.}\binits{J.}},
  \bauthor{\bsnm{Pacaud},~\bfnm{F.}\binits{F.}},
  \bauthor{\bsnm{R{\'e}fr{\'e}gier},~\bfnm{A.}\binits{A.}},
  \bauthor{\bsnm{Massey},~\bfnm{R.}\binits{R.}},
  \bauthor{\bsnm{Pierre},~\bfnm{M.}\binits{M.}},
  \bauthor{\bsnm{Amara},~\bfnm{A.}\binits{A.}},
  \bauthor{\bsnm{Birkinshaw},~\bfnm{M.}\binits{M.}},
  \bauthor{\bsnm{Paulin-Henriksson},~\bfnm{S.}\binits{S.}},
  \bauthor{\bsnm{Smith},~\bfnm{G.~P.}\binits{G.~P.}} \AND
  \bauthor{\bsnm{Willis},~\bfnm{J.}\binits{J.}}
(\byear{2008}).
\btitle{Combined analysis of weak lensing and X-ray blind surveys}.
\bjournal{Monthly Notices of the RAS}
\bvolume{385}
\bpages{695--707}.
\end{barticle}
\endbibitem

\bibitem[\protect\citeauthoryear{Bernstein}{2010}]{2010arXiv10012333B}
\begin{bmisc}[author]
\bauthor{\bsnm{Bernstein},~\bfnm{G.~M.}\binits{G.~M.}}
(\byear{2010}).
\bhowpublished{Shape measurement biases from underfitting and ellipticity gradients.
Available at \url{http://arxiv.org/abs/1001.2333}}.
\end{bmisc}
\endbibitem

\bibitem[\protect\citeauthoryear{Bernstein and Jarvis}{2002}]{2002AJ123583B}
\begin{barticle}[author]
\bauthor{\bsnm{Bernstein},~\bfnm{G.~M.}\binits{G.~M.}} \AND
  \bauthor{\bsnm{Jarvis},~\bfnm{M.}\binits{M.}}
(\byear{2002}).
\btitle{Shapes and shears, stars and smears: Optimal measurements for weak
  lensing}.
\bjournal{Astrophysics Journal}
\bvolume{123}
\bpages{583--618}.
\end{barticle}
\endbibitem

\bibitem[\protect\citeauthoryear{Bridle et al.}{2009}]{2009AnApS36B}
\begin{barticle}[author]
\bauthor{\bsnm{Bridle},~\bfnm{S.}\binits{S.}},
  \bauthor{\bsnm{Shawe-Taylor},~\bfnm{J.}\binits{J.}},
  \bauthor{\bsnm{Amara},~\bfnm{A.}\binits{A.}},
  \bauthor{\bsnm{Applegate},~\bfnm{D.}\binits{D.}},
  \bauthor{\bsnm{Balan},~\bfnm{Joel S.~T.}\binits{J.~S.~T.} \bsuffix{Berge}},
  \bauthor{\bsnm{Bernstein},~\bfnm{G.}\binits{G.}},
  \bauthor{\bsnm{Dahle},~\bfnm{H.}\binits{H.}},
  \bauthor{\bsnm{Erben},~\bfnm{T.}\binits{T.}},
  \bauthor{\bsnm{Gill},~\bfnm{M.}\binits{M.}},
  \bauthor{\bsnm{Heavens},~\bfnm{A.}\binits{A.}},
  \bauthor{\bsnm{Heymans},~\bfnm{C.}\binits{C.}},
  \bauthor{\bsnm{High},~\bfnm{F.~W.}\binits{F.~W.}},
  \bauthor{\bsnm{Hoekstra},~\bfnm{H.}\binits{H.}},
  \bauthor{\bsnm{Jarvis},~\bfnm{M.}\binits{M.}},
  \bauthor{\bsnm{Kirk},~\bfnm{D.}\binits{D.}},
  \bauthor{\bsnm{Kitching},~\bfnm{T.}\binits{T.}},
  \bauthor{\bsnm{Kneib},~\bfnm{J.~P.}\binits{J.~P.}},
  \bauthor{\bsnm{Kuijken},~\bfnm{K.}\binits{K.}},
  \bauthor{\bsnm{Lagatutta},~\bfnm{D.}\binits{D.}},
  \bauthor{\bsnm{Mandelbaum},~\bfnm{R.}\binits{R.}},
  \bauthor{\bsnm{Massey},~\bfnm{R.}\binits{R.}},
  \bauthor{\bsnm{Mellier},~\bfnm{Y.}\binits{Y.}},
  \bauthor{\bsnm{Moghaddam},~\bfnm{B.}\binits{B.}},
  \bauthor{\bsnm{Moudden},~\bfnm{Y.}\binits{Y.}},
  \bauthor{\bsnm{Nakajima},~\bfnm{R.}\binits{R.}},
  \bauthor{\bsnm{Paulin-Henriksson},~\bfnm{S.}\binits{S.}},
  \bauthor{\bsnm{Pires},~\bfnm{S.}\binits{S.}},
  \bauthor{\bsnm{Rassat},~\bfnm{A.}\binits{A.}},
  \bauthor{\bsnm{Refregier},~\bfnm{A.}\binits{A.}},
  \bauthor{\bsnm{Rhodes},~\bfnm{J.}\binits{J.}},
  \bauthor{\bsnm{Schrabback},~\bfnm{T.}\binits{T.}},
  \bauthor{\bsnm{Semboloni},~\bfnm{E.}\binits{E.}},
  \bauthor{\bsnm{Shmakova},~\bfnm{M.}\binits{M.}}, \bauthor{\bparticle{van}
  \bsnm{Waerbeke},~\bfnm{L.}\binits{L.}},
  \bauthor{\bsnm{Witherick},~\bfnm{D.}\binits{D.}},
  \bauthor{\bsnm{Voigt},~\bfnm{L.}\binits{L.}} \AND
  \bauthor{\bsnm{Wittman},~\bfnm{D.}\binits{D.}}
(\byear{2009}).
\btitle{Handbook for the GREAT08 Challenge: An image analysis competition for
  cosmological lensing}.
\bjournal{Ann. Appl. Stat.}
\bvolume{3}
\bpages{6--37}.
\end{barticle}
\MR{2668698}
\endbibitem


\bibitem[\protect\citeauthoryear{Cho and Lee}{2009}]{Cho2009}
\begin{barticle}[author]
\bauthor{\bsnm{Cho},~\bfnm{S.}\binits{S.}} \AND
  \bauthor{\bsnm{Lee},~\bfnm{S.}\binits{S.}}
(\byear{2009}).
\btitle{Fast Motion Deblurring}.
\bjournal{ACM Transactions on Graphics (SIGGRAPH ASIA 2009)}
\bvolume{28}
\bpages{Art. 145}.
\end{barticle}
\endbibitem

\bibitem[\protect\citeauthoryear{Cressie}{1991}]{Cressie}
\begin{bbook}[author]
\bauthor{\bsnm{Cressie},~\bfnm{N.}\binits{N.}}
(\byear{1991}).
\btitle{Statistics for Spatial Data}.
\bpublisher{Wiley}, \baddress{New York}.
\end{bbook}
\MR{1239641}
\endbibitem


\bibitem[\protect\citeauthoryear{Crittenden et al.}{2002}]{2002ApJ56820C}
\begin{barticle}[author]
\bauthor{\bsnm{Crittenden},~\bfnm{R.~G.}\binits{R.~G.}},
  \bauthor{\bsnm{Natarajan},~\bfnm{P.}\binits{P.}},
  \bauthor{\bsnm{Pen},~\bfnm{U.~L.}\binits{U.~L.}} \AND
  \bauthor{\bsnm{Theuns},~\bfnm{T.}\binits{T.}}
(\byear{2002}).
\btitle{Discriminating weak lensing from intrinsic spin correlations using the
  curl-gradient decomposition}.
\bjournal{Astrophysical J.}
\bvolume{568}
\bpages{20--27}.
\end{barticle}
\endbibitem

\bibitem[\protect\citeauthoryear{Fergus et al.}{2006}]{Fergus2006}
\begin{bmisc}[author]
\bauthor{\bsnm{Fergus},~\bfnm{R.}\binits{R.}},
  \bauthor{\bsnm{Singh},~\bfnm{B.}\binits{B.}},
  \bauthor{\bsnm{Hertzmann},~\bfnm{A.}\binits{A.}},
  \bauthor{\bsnm{Roweis},~\bfnm{S.~T.}\binits{S.~T.}} \AND
  \bauthor{\bsnm{Freeman},~\bfnm{W.~T.}\binits{W.~T.}}
(\byear{2006}).
\bhowpublished{Removing camera shake from a single image.
In \textit{ACM Transactions on Graphics} (\textit{SIGGRAPH}). ACM.}
\end{bmisc}
\endbibitem

\bibitem[\protect\citeauthoryear{Fu et al.}{2008}]{2008AA4799F}
\begin{barticle}[author]
\bauthor{\bsnm{Fu},~\bfnm{L.}\binits{L.}},
  \bauthor{\bsnm{Semboloni},~\bfnm{E.}\binits{E.}},
  \bauthor{\bsnm{Hoekstra},~\bfnm{H.}\binits{H.}},
  \bauthor{\bsnm{Kilbinger},~\bfnm{M.}\binits{M.}}, \bauthor{\bparticle{van}
  \bsnm{Waerbeke},~\bfnm{L.}\binits{L.}},
  \bauthor{\bsnm{Tereno},~\bfnm{I.}\binits{I.}},
  \bauthor{\bsnm{Mellier},~\bfnm{Y.}\binits{Y.}},
  \bauthor{\bsnm{Heymans},~\bfnm{C.}\binits{C.}},
  \bauthor{\bsnm{Coupon},~\bfnm{J.}\binits{J.}},
  \bauthor{\bsnm{Benabed},~\bfnm{K.}\binits{K.}},
  \bauthor{\bsnm{Benjamin},~\bfnm{J.}\binits{J.}},
  \bauthor{\bsnm{Bertin},~\bfnm{E.}\binits{E.}},
  \bauthor{\bsnm{Dor{\'e}},~\bfnm{O.}\binits{O.}},
  \bauthor{\bsnm{Hudson},~\bfnm{M.~J.}\binits{M.~J.}},
  \bauthor{\bsnm{Ilbert},~\bfnm{O.}\binits{O.}},
  \bauthor{\bsnm{Maoli},~\bfnm{R.}\binits{R.}},
  \bauthor{\bsnm{Marmo},~\bfnm{C.}\binits{C.}},
  \bauthor{\bsnm{McCracken},~\bfnm{H.~J.}\binits{H.~J.}} \AND
  \bauthor{\bsnm{M{\'e}nard},~\bfnm{B.}\binits{B.}}
(\byear{2008}).
\btitle{Very weak lensing in the CFHTLS wide: Cosmology from cosmic shear in
  the linear regime}.
\bjournal{Astronomy and Astrophysics Proceedings}
\bvolume{479}
\bpages{9--25}.
\end{barticle}
\endbibitem

\bibitem[\protect\citeauthoryear{Goldberg and Bacon}{2005}]{2005ApJ619741G}
\begin{barticle}[author]
\bauthor{\bsnm{Goldberg},~\bfnm{D.~M.}\binits{D.~M.}} \AND
  \bauthor{\bsnm{Bacon},~\bfnm{D.~J.}\binits{D.~J.}}
(\byear{2005}).
\btitle{Galaxy--galaxy flexion: Weak lensing to second order}.
\bjournal{Astrophysical J.}
\bvolume{619}
\bpages{741--748}.
\end{barticle}
\endbibitem

\bibitem[\protect\citeauthoryear{Gruen et al.}{2010}]{2010arXiv10020838G}
\begin{bmisc}[author]
\bauthor{\bsnm{Gruen},~\bfnm{D.}\binits{D.}},
  \bauthor{\bsnm{Seitz},~\bfnm{S.}\binits{S.}},
  \bauthor{\bsnm{Koppenhoefer},~\bfnm{J.}\binits{J.}} \AND
  \bauthor{\bsnm{Riffeser},~\bfnm{A.}\binits{A.}}
(\byear{2010}).
\bhowpublished{Bias-free shear estimation using artificial neural networks.
Available at \texttt{\href{http://arxiv.org/abs/1002.0838}{http://arxiv.org/abs/}
\href{http://arxiv.org/abs/1002.0838}{1002.0838}}.}
\end{bmisc}
\endbibitem

\bibitem[\protect\citeauthoryear{Harmeling et al.}{2009}]{harmeling09}
\begin{binproceedings}[author]
\bauthor{\bsnm{Harmeling},~\bfnm{S.}\binits{S.}},
  \bauthor{\bsnm{Hirsch},~\bfnm{M.}\binits{M.}},
  \bauthor{\bsnm{Sra},~\bfnm{S.}\binits{S.}} \AND
  \bauthor{\bsnm{Sch{\"o}lkopf},~\bfnm{B.}\binits{B.}}
(\byear{2009}).
\btitle{Online blind image deconvolution for astronomy}.
In \bbooktitle{Proceedings of the IEEE Conference on Comp.~Photogr.}
\end{binproceedings}
\endbibitem

\bibitem[\protect\citeauthoryear{Heymans et al.}{2006}]{2006MNRAS3681323H}
\begin{barticle}[author]
\bauthor{\bsnm{Heymans},~\bfnm{C.}\binits{C.}},
  \bauthor{\bsnm{Van~Waerbeke},~\bfnm{L.}\binits{L.}},
  \bauthor{\bsnm{Bacon},~\bfnm{D.}\binits{D.}},
  \bauthor{\bsnm{Berge},~\bfnm{J.}\binits{J.}},
  \bauthor{\bsnm{Bernstein},~\bfnm{G.}\binits{G.}},
  \bauthor{\bsnm{Bertin},~\bfnm{E.}\binits{E.}},
  \bauthor{\bsnm{Bridle},~\bfnm{S.}\binits{S.}},
  \bauthor{\bsnm{Brown},~\bfnm{M.~L.}\binits{M.~L.}},
  \bauthor{\bsnm{Clowe},~\bfnm{D.}\binits{D.}},
  \bauthor{\bsnm{Dahle},~\bfnm{H.}\binits{H.}},
  \bauthor{\bsnm{Erben},~\bfnm{T.}\binits{T.}},
  \bauthor{\bsnm{Gray},~\bfnm{M.}\binits{M.}},
  \bauthor{\bsnm{Hetterscheidt},~\bfnm{M.}\binits{M.}},
  \bauthor{\bsnm{Hoekstra},~\bfnm{H.}\binits{H.}},
  \bauthor{\bsnm{Hudelot},~\bfnm{P.}\binits{P.}},
  \bauthor{\bsnm{Jarvis},~\bfnm{M.}\binits{M.}},
  \bauthor{\bsnm{Kuijken},~\bfnm{K.}\binits{K.}},
  \bauthor{\bsnm{Margoniner},~\bfnm{V.}\binits{V.}},
  \bauthor{\bsnm{Massey},~\bfnm{R.}\binits{R.}},
  \bauthor{\bsnm{Mellier},~\bfnm{Y.}\binits{Y.}},
  \bauthor{\bsnm{Nakajima},~\bfnm{R.}\binits{R.}},
  \bauthor{\bsnm{Refregier},~\bfnm{A.}\binits{A.}},
  \bauthor{\bsnm{Rhodes},~\bfnm{J.}\binits{J.}},
  \bauthor{\bsnm{Schrabback},~\bfnm{T.}\binits{T.}} \AND
  \bauthor{\bsnm{Wittman},~\bfnm{D.}\binits{D.}}
(\byear{2006}).
\btitle{The Shear Testing Programme---I. Weak lensing analysis of simulated
  ground-based observations}.
\bjournal{Monthly Notices of the RAS}
\bvolume{368}
\bpages{1323--1339}.
\end{barticle}
\endbibitem

\bibitem[\protect\citeauthoryear{Hirsch et al.}{2010}]{hirsch10}
\begin{binproceedings}[author]
\bauthor{\bsnm{Hirsch},~\bfnm{M.}\binits{M.}},
  \bauthor{\bsnm{Sra},~\bfnm{S.}\binits{S.}},
  \bauthor{\bsnm{Sch{\"o}lkopf},~\bfnm{B.}\binits{B.}} \AND
  \bauthor{\bsnm{Harmeling},~\bfnm{S.}\binits{S.}}
(\byear{2010}).
\btitle{Efficient filter flow for space-variant multiframe blind
  deconvolution}.
In \bbooktitle{IEEE Computer Vision and Pattern Recognition}.
\end{binproceedings}
\endbibitem

\bibitem[\protect\citeauthoryear{Hosseini and Bethge}{2009}]{Hosseini}
\begin{bmisc}[author]
\bauthor{\bsnm{Hosseini},~\bfnm{R.}\binits{R.}} \AND
\bauthor{\bsnm{Bethge},~\bfnm{M.}\binits{M.}}
(\byear{2009}).
\bhowpublished{Max planck institute for biological cybernetics technical report}.
\end{bmisc}
\endbibitem

\bibitem[\protect\citeauthoryear{Jarvis and Jain}{2004}]{2004astroph12234J}
\begin{bmisc}[author]
\bauthor{\bsnm{Jarvis},~\bfnm{M.}\binits{M.}} \AND
  \bauthor{\bsnm{Jain},~\bfnm{B.}\binits{B.}}
(\byear{2004}).
\bhowpublished{Principal component analysis of PSF variation in weak lensing surveys.
Available at \url{http://arxiv.org/abs/astro-ph/0412234}.}
\end{bmisc}
\endbibitem

\bibitem[\protect\citeauthoryear{Jarvis, Schechter and
  Jain}{2008}]{2008arXiv08100027J}
\begin{bmisc}[author]
\bauthor{\bsnm{Jarvis},~\bfnm{M.}\binits{M.}},
  \bauthor{\bsnm{Schechter},~\bfnm{P.}\binits{P.}} \AND
  \bauthor{\bsnm{Jain},~\bfnm{B.}\binits{B.}}
(\byear{2008}).
\bhowpublished{Telescope optics and weak lensing: PSF patterns due to low order
  aberrations.
Available at \url{http://arxiv.org/abs/0810.0027}.}
\end{bmisc}
\endbibitem


\bibitem[\protect\citeauthoryear{Kalirai et al.}{2003}]{2003AJ1261402K}
\begin{barticle}[author]
\bauthor{\bsnm{Kalirai},~\bfnm{J.~S.}\binits{J.~S.}},
  \bauthor{\bsnm{Fahlman},~\bfnm{G.~G.}\binits{G.~G.}},
  \bauthor{\bsnm{Richer},~\bfnm{H.~B.}\binits{H.~B.}} \AND
  \bauthor{\bsnm{Ventura},~\bfnm{P.}\binits{P.}}
(\byear{2003}).
\btitle{The CFHT open star cluster survey. IV. Two rich, young open star
  clusters: NGC 2168 (M35) and NGC 2323 (M50)}.
\bjournal{Astrophysics Journal}
\bvolume{126}
\bpages{1402--1414}.
\end{barticle}
\endbibitem


\bibitem[\protect\citeauthoryear{Kitching et al.}{2008}]{2008MNRAS390149K}
\begin{barticle}[author]
\bauthor{\bsnm{Kitching},~\bfnm{T.~D.}\binits{T.~D.}},
  \bauthor{\bsnm{Miller},~\bfnm{L.}\binits{L.}},
  \bauthor{\bsnm{Heymans},~\bfnm{C.~E.}\binits{C.~E.}},
  \bauthor{\bparticle{van} \bsnm{Waerbeke},~\bfnm{L.}\binits{L.}} \AND
  \bauthor{\bsnm{Heavens},~\bfnm{A.~F.}\binits{A.~F.}}
(\byear{2008}).
\btitle{Bayesian galaxy shape measurement for weak lensing surveys---II.
  Application to simulations}.
\bjournal{Monthly Notices of the RAS}
\bvolume{390}
\bpages{149--167}.
\end{barticle}
\endbibitem

\bibitem[\protect\citeauthoryear{Knox and Thompson}{1974}]{knox}
\begin{barticle}[author]
\bauthor{\bsnm{Knox},~\bfnm{K.~T.}\binits{K.~T.}} \AND
  \bauthor{\bsnm{Thompson},~\bfnm{B.~J.}\binits{B.~J.}}
(\byear{1974}).
\btitle{Recovery of images from atmospherically degraded short-exposure
  photographs}.
\bjournal{Astrophysical J.}
\bvolume{193}
\bpages{L45--L48}.
\end{barticle}
\endbibitem

\bibitem[\protect\citeauthoryear{Kuijken}{2006}]{2006astroph10606K}
\begin{bmisc}[author]
\bauthor{\bsnm{Kuijken},~\bfnm{K.}\binits{K.}}
(\byear{2006}).
\bhowpublished{GaaP: PSF- and aperture-matched photometry using shapelets.
Available at \url{http://arxiv.org/abs/astro-ph/0610606}.}
\end{bmisc}
\endbibitem

\bibitem[\protect\citeauthoryear{Kundur and Hatzinakos}{1996}]{kundur}
\begin{barticle}[author]
\bauthor{\bsnm{Kundur},~\bfnm{D.}\binits{D.}} \AND
  \bauthor{\bsnm{Hatzinakos},~\bfnm{D.}\binits{D.}}
(\byear{1996}).
\btitle{Blind image deconvolution}.
\bjournal{IEEE Signal Processing Mag.}
\bvolume{13}
\bpages{43--64}.
\end{barticle}
\endbibitem

\bibitem[\protect\citeauthoryear{Labeyrie}{1970}]{labeyrie}
\begin{barticle}[author]
\bauthor{\bsnm{Labeyrie},~\bfnm{A.}\binits{A.}}
(\byear{1970}).
\btitle{Attainment of diffraction limited resolution in large telescopes by
  fourier analysing speckle patterns in star images}.
\bjournal{Astron. Astrophys.}
\bvolume{6}
\bpages{85--87}.
\end{barticle}
\endbibitem

\bibitem[\protect\citeauthoryear{Massey and Refregier}{2005}]{2005MNRAS363197M}
\begin{barticle}[author]
\bauthor{\bsnm{Massey},~\bfnm{R.}\binits{R.}} \AND
  \bauthor{\bsnm{Refregier},~\bfnm{A.}\binits{A.}}
(\byear{2005}).
\btitle{Polar shapelets}.
\bjournal{Monthly Notices of the RAS}
\bvolume{363}
\bpages{197--210}.
\end{barticle}
\endbibitem

\bibitem[\protect\citeauthoryear{Massey et al.}{2007}]{2007MNRAS37613M}
\begin{barticle}[author]
\bauthor{\bsnm{Massey},~\bfnm{R.}\binits{R.}},
  \bauthor{\bsnm{Heymans},~\bfnm{C.}\binits{C.}},
  \bauthor{\bsnm{Berg{\'e}},~\bfnm{J.}\binits{J.}},
  \bauthor{\bsnm{Bernstein},~\bfnm{G.}\binits{G.}},
  \bauthor{\bsnm{Bridle},~\bfnm{S.}\binits{S.}},
  \bauthor{\bsnm{Clowe},~\bfnm{D.}\binits{D.}},
  \bauthor{\bsnm{Dahle},~\bfnm{H.}\binits{H.}},
  \bauthor{\bsnm{Ellis},~\bfnm{R.}\binits{R.}},
  \bauthor{\bsnm{Erben},~\bfnm{T.}\binits{T.}},
  \bauthor{\bsnm{Hetterscheidt},~\bfnm{M.}\binits{M.}},
  \bauthor{\bsnm{High},~\bfnm{F.~W.}\binits{F.~W.}},
  \bauthor{\bsnm{Hirata},~\bfnm{C.}\binits{C.}},
  \bauthor{\bsnm{Hoekstra},~\bfnm{H.}\binits{H.}},
  \bauthor{\bsnm{Hudelot},~\bfnm{P.}\binits{P.}},
  \bauthor{\bsnm{Jarvis},~\bfnm{M.}\binits{M.}},
  \bauthor{\bsnm{Johnston},~\bfnm{D.}\binits{D.}},
  \bauthor{\bsnm{Kuijken},~\bfnm{K.}\binits{K.}},
  \bauthor{\bsnm{Margoniner},~\bfnm{V.}\binits{V.}},
  \bauthor{\bsnm{Mandelbaum},~\bfnm{R.}\binits{R.}},
  \bauthor{\bsnm{Mellier},~\bfnm{Y.}\binits{Y.}},
  \bauthor{\bsnm{Nakajima},~\bfnm{R.}\binits{R.}},
  \bauthor{\bsnm{Paulin-Henriksson},~\bfnm{S.}\binits{S.}},
  \bauthor{\bsnm{Peeples},~\bfnm{M.}\binits{M.}},
  \bauthor{\bsnm{Roat},~\bfnm{C.}\binits{C.}},
  \bauthor{\bsnm{Refregier},~\bfnm{A.}\binits{A.}},
  \bauthor{\bsnm{Rhodes},~\bfnm{J.}\binits{J.}},
  \bauthor{\bsnm{Schrabback},~\bfnm{T.}\binits{T.}},
  \bauthor{\bsnm{Schirmer},~\bfnm{M.}\binits{M.}},
  \bauthor{\bsnm{Seljak},~\bfnm{U.}\binits{U.}},
  \bauthor{\bsnm{Semboloni},~\bfnm{E.}\binits{E.}} \AND
  \bauthor{\bparticle{van} \bsnm{Waerbeke},~\bfnm{L.}\binits{L.}}
(\byear{2007}).
\btitle{The shear testing programme 2: Factors affecting high-precision
  weak-lensing analyses}.
\bjournal{Monthly Notices of the RAS}
\bvolume{376}
\bpages{13--38}.
\end{barticle}
\endbibitem

\bibitem[\protect\citeauthoryear{Melchior et al.}{2010}]{2010arXiv10081076M}
\begin{bmisc}[author]
\bauthor{\bsnm{Melchior},~\bfnm{P.}\binits{P.}},
  \bauthor{\bsnm{Viola},~\bfnm{M.}\binits{M.}},
  \bauthor{\bsnm{Sch{\"a}fer},~\bfnm{B.~M.}\binits{B.~M.}} \AND
  \bauthor{\bsnm{Bartelmann},~\bfnm{M.}\binits{M.}}
(\byear{2010}).
\bhowpublished{Weak gravitational lensing with DEIMOS.
Available at
\texttt{\href{http://adsabs.harvard.edu/abs/2010arXiv1008.1076M}{http://adsabs.harvard.edu/abs/}
\href{http://adsabs.harvard.edu/abs/2010arXiv1008.1076M}{2010arXiv1008.1076M}}.}
\end{bmisc}
\endbibitem

\bibitem[\protect\citeauthoryear{Miller et al.}{2007}]{2007MNRAS382315M}
\begin{barticle}[author]
\bauthor{\bsnm{Miller},~\bfnm{L.}\binits{L.}},
  \bauthor{\bsnm{Kitching},~\bfnm{T.~D.}\binits{T.~D.}},
  \bauthor{\bsnm{Heymans},~\bfnm{C.}\binits{C.}},
  \bauthor{\bsnm{Heavens},~\bfnm{A.~F.}\binits{A.~F.}} \AND
  \bauthor{\bparticle{van} \bsnm{Waerbeke},~\bfnm{L.}\binits{L.}}
(\byear{2007}).
\btitle{Bayesian galaxy shape measurement for weak lensing surveys---I.
  Methodology and a fast-fitting algorithm}.
\bjournal{Monthly Notices of the RAS}
\bvolume{382}
\bpages{315--324}.
\end{barticle}
\endbibitem

\bibitem[\protect\citeauthoryear{Nagy and O'Leary}{1998}]{Nagy1998}
\begin{barticle}[author]
\bauthor{\bsnm{Nagy},~\bfnm{J.~G.}\binits{J.~G.}} \AND
  \bauthor{\bsnm{O'Leary},~\bfnm{D.~P.}\binits{D.~P.}}
(\byear{1998}).
\btitle{Restoring images degraded by spatially variant blur}.
\bjournal{SIAM J. Sci. Comput.}
\bvolume{19}
\bpages{1063--1082}.
\end{barticle}
\MR{1614295}
\endbibitem

\bibitem[\protect\citeauthoryear{Paulin-Henriksson, Refregier and
  Amara}{2009}]{2009AA500647P}
\begin{barticle}[author]
\bauthor{\bsnm{Paulin-Henriksson},~\bfnm{S.}\binits{S.}},
  \bauthor{\bsnm{Refregier},~\bfnm{A.}\binits{A.}} \AND
  \bauthor{\bsnm{Amara},~\bfnm{A.}\binits{A.}}
(\byear{2009}).
\btitle{Optimal point spread function modeling for weak lensing: Complexity and
  sparsity}.
\bjournal{Astronomy and Astrophysics Proceedings}
\bvolume{500}
\bpages{647--655}.
\end{barticle}
\endbibitem

\bibitem[\protect\citeauthoryear{Paulin-Henriksson et al.}{2008}]{2008AA48467P}
\begin{barticle}[author]
\bauthor{\bsnm{Paulin-Henriksson},~\bfnm{S.}\binits{S.}},
  \bauthor{\bsnm{Amara},~\bfnm{A.}\binits{A.}},
  \bauthor{\bsnm{Voigt},~\bfnm{L.}\binits{L.}},
  \bauthor{\bsnm{Refregier},~\bfnm{A.}\binits{A.}} \AND
  \bauthor{\bsnm{Bridle},~\bfnm{S.~L.}\binits{S.~L.}}
(\byear{2008}).
\btitle{Point spread function calibration requirements for dark energy from
  cosmic shear}.
\bjournal{Astronomy and Astrophysics Proceedings}
\bvolume{484}
\bpages{67--77}.
\end{barticle}
\endbibitem

\bibitem[\protect\citeauthoryear{Refregier}{2003}]{2003ARAA41645R}
\begin{barticle}[author]
\bauthor{\bsnm{Refregier},~\bfnm{A.}\binits{A.}}
(\byear{2003}).
\btitle{Weak gravitational lensing by large-scale structure}.
\bjournal{Araa}
\bvolume{41}
\bpages{645--668}.
\end{barticle}
\endbibitem


\bibitem[\protect\citeauthoryear{Rhodes et al.}{2007}]{2007ApJS172203R}
\begin{barticle}[author]
\bauthor{\bsnm{Rhodes},~\bfnm{J.~D.}\binits{J.~D.}},
  \bauthor{\bsnm{Massey},~\bfnm{R.~J.}\binits{R.~J.}},
  \bauthor{\bsnm{Albert},~\bfnm{J.}\binits{J.}},
  \bauthor{\bsnm{Collins},~\bfnm{N.}\binits{N.}},
  \bauthor{\bsnm{Ellis},~\bfnm{R.~S.}\binits{R.~S.}},
  \bauthor{\bsnm{Heymans},~\bfnm{C.}\binits{C.}},
  \bauthor{\bsnm{Gardner},~\bfnm{J.~P.}\binits{J.~P.}},
  \bauthor{\bsnm{Kneib},~\bfnm{J.~P.}\binits{J.~P.}},
  \bauthor{\bsnm{Koekemoer},~\bfnm{A.}\binits{A.}},
  \bauthor{\bsnm{Leauthaud},~\bfnm{A.}\binits{A.}},
  \bauthor{\bsnm{Mellier},~\bfnm{Y.}\binits{Y.}},
  \bauthor{\bsnm{Refregier},~\bfnm{A.}\binits{A.}},
  \bauthor{\bsnm{Taylor},~\bfnm{J.~E.}\binits{J.~E.}} \AND
  \bauthor{\bsnm{Van~Waerbeke},~\bfnm{L.}\binits{L.}}
(\byear{2007}).
\btitle{The stability of the point-spread function of the advanced camera for
  surveys on the hubble space telescope and implications for weak gravitational
  lensing}.
\bjournal{Astrophysical J. Supplement}
\bvolume{172}
\bpages{203--218}.
\end{barticle}
\endbibitem

\bibitem[\protect\citeauthoryear{Rowe}{2010}]{2010MNRAS404350R}
\begin{barticle}[author]
\bauthor{\bsnm{Rowe},~\bfnm{B.}\binits{B.}}
(\byear{2010}).
\btitle{Improving PSF modelling for weak gravitational lensing using new
  methods in model selection}.
\bjournal{Monthly Notices of the RAS}
\bvolume{404}
\bpages{350--366}.
\end{barticle}
\endbibitem

\bibitem[\protect\citeauthoryear{Schneider et al.}{2002}]{2002AA3961S}
\begin{barticle}[author]
\bauthor{\bsnm{Schneider},~\bfnm{P.}\binits{P.}}, \bauthor{\bparticle{van}
  \bsnm{Waerbeke},~\bfnm{L.}\binits{L.}},
  \bauthor{\bsnm{Kilbinger},~\bfnm{M.}\binits{M.}} \AND
  \bauthor{\bsnm{Mellier},~\bfnm{Y.}\binits{Y.}}
(\byear{2002}).
\btitle{Analysis of two-point statistics of cosmic shear. I. Estimators and
  covariances}.
\bjournal{Astronomy and Astrophysics Proceedings}
\bvolume{396}
\bpages{1--19}.
\end{barticle}
\endbibitem

\bibitem[\protect\citeauthoryear{Sch{\"o}lkopf, Smola and
  M{\"u}ller}{1998}]{SSM98}
\begin{barticle}[author]
\bauthor{\bsnm{Sch{\"o}lkopf},~\bfnm{Bernhard}\binits{B.}},
  \bauthor{\bsnm{Smola},~\bfnm{Alexander}\binits{A.}} \AND
  \bauthor{\bsnm{M{\"u}ller},~\bfnm{Klaus-Robert}\binits{K.-R.}}
(\byear{1998}).
\btitle{Nonlinear component analysis as a kernel eigenvalue problem}.
\bjournal{Neural Comput.}
\bvolume{10}
\bpages{1299--1319}.
\end{barticle}
\endbibitem

\bibitem[\protect\citeauthoryear{Schrabback et al.}{2010}]{2010AA516A63S}
\begin{barticle}[author]
\bauthor{\bsnm{Schrabback},~\bfnm{T.}\binits{T.}},
  \bauthor{\bsnm{Hartlap},~\bfnm{J.}\binits{J.}},
  \bauthor{\bsnm{Joachimi},~\bfnm{B.}\binits{B.}},
  \bauthor{\bsnm{Kilbinger},~\bfnm{M.}\binits{M.}},
  \bauthor{\bsnm{Simon},~\bfnm{P.}\binits{P.}},
  \bauthor{\bsnm{Benabed},~\bfnm{K.}\binits{K.}}, \bauthor{\bsnm{Brada{\v
  c}},~\bfnm{M.}\binits{M.}}, \bauthor{\bsnm{Eifler},~\bfnm{T.}\binits{T.}},
  \bauthor{\bsnm{Erben},~\bfnm{T.}\binits{T.}},
  \bauthor{\bsnm{Fassnacht},~\bfnm{C.~D.}\binits{C.~D.}},
  \bauthor{\bsnm{High},~\bfnm{F.~W.}\binits{F.~W.}},
  \bauthor{\bsnm{Hilbert},~\bfnm{S.}\binits{S.}},
  \bauthor{\bsnm{Hildebrandt},~\bfnm{H.}\binits{H.}},
  \bauthor{\bsnm{Hoekstra},~\bfnm{H.}\binits{H.}},
  \bauthor{\bsnm{Kuijken},~\bfnm{K.}\binits{K.}},
  \bauthor{\bsnm{Marshall},~\bfnm{P.~J.}\binits{P.~J.}},
  \bauthor{\bsnm{Mellier},~\bfnm{Y.}\binits{Y.}},
  \bauthor{\bsnm{Morganson},~\bfnm{E.}\binits{E.}},
  \bauthor{\bsnm{Schneider},~\bfnm{P.}\binits{P.}},
  \bauthor{\bsnm{Semboloni},~\bfnm{E.}\binits{E.}}, \bauthor{\bparticle{van}
  \bsnm{Waerbeke},~\bfnm{L.}\binits{L.}} \AND
  \bauthor{\bsnm{Velander},~\bfnm{M.}\binits{M.}}
(\byear{2010}).
\btitle{Evidence of the accelerated expansion of the Universe from weak lensing
  tomography with COSMOS}.
\bjournal{Astronomy and Astrophysics Proceedings}
\bvolume{516}
\bpages{A63+}.
\end{barticle}
\endbibitem

\bibitem[\protect\citeauthoryear{Schulz}{1993}]{schulz}
\begin{barticle}[author]
\bauthor{\bsnm{Schulz},~\bfnm{T.~J.}\binits{T.~J.}}
(\byear{1993}).
\btitle{Multiframe blind deconvolution of astronomical images.}
\bjournal{J. Opt. Soc. Amer. A}
\bvolume{10}
\bpages{1064--1073}.
\end{barticle}
\endbibitem

\bibitem[\protect\citeauthoryear{Shawe-Taylor and Cristianini}{2004}]{ecs9580}
\begin{bbook}[author]
\bauthor{\bsnm{Shawe-Taylor},~\bfnm{J.}\binits{J.}} \AND
  \bauthor{\bsnm{Cristianini},~\bfnm{N.}\binits{N.}}
(\byear{2004}).
\btitle{Kernel Methods for Pattern Analysis}.
\bpublisher{Cambridge Univ. Press}, \baddress{Cambridge}.
\end{bbook}
\endbibitem


\bibitem[\protect\citeauthoryear{Van~Waerbeke, Mellier and
  Hoekstra}{2005}]{2005AA42975V}
\begin{barticle}[author]
\bauthor{\bsnm{Van~Waerbeke},~\bfnm{L.}\binits{L.}},
  \bauthor{\bsnm{Mellier},~\bfnm{Y.}\binits{Y.}} \AND
  \bauthor{\bsnm{Hoekstra},~\bfnm{H.}\binits{H.}}
(\byear{2005}).
\btitle{Dealing with systematics in cosmic shear studies: New results from the
  VIRMOS-Descart survey}.
\bjournal{Astronomy and Astrophysics Proceedings}
\bvolume{429}
\bpages{75--84}.
\end{barticle}
\endbibitem

\bibitem[\protect\citeauthoryear{Zhang}{2010}]{2010arXiv10023615Z}
\begin{bmisc}[author]
\bauthor{\bsnm{Zhang},~\bfnm{J.}\binits{J.}}
(\byear{2010}).
\bhowpublished{Ideal\vspace*{-1pt} cosmic shear estimators do not exist.
Available at \texttt{\href{http://arxiv.org/abs/1002.3615}{http://}
\href{http://arxiv.org/abs/1002.3615}{arxiv.org/abs/1002.3615}}.}
\end{bmisc}
\endbibitem

\end{thebibliography}
\end{document}